\documentclass{aa}  

\usepackage{ulem}
\usepackage{graphicx}
\usepackage{txfonts}
\usepackage{amsmath}
\usepackage{amssymb}
\usepackage{subcaption}
\usepackage{multirow}
\usepackage{adjustbox}
\usepackage{comment}
\usepackage{tabto}

\usepackage[colorlinks,breaklinks]{hyperref}
\hypersetup{linkcolor=blue,citecolor=blue,filecolor=black,urlcolor=blue}

\usepackage[switch]{lineno}

\newcommand{\orcid}[1]{\href{https://orcid.org/#1}{\includegraphics[height=11pt]{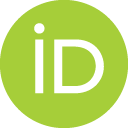}}}

\defcitealias{Childress2014}{C14}
\defcitealias{Nicolas2021}{N21}
\defcitealias{Rose2019}{R19}

\begin{document} 

\title{Which Type Ia supernova observables best indicate the ages of their progenitor stars? }
\titlerunning{SN Ia's Progenitor Age and $x_{1}$ and $c$}
\authorrunning{Y.-L. Kim et al.}

\author{
	Young-Lo~Kim\thanks{ylkim83@yonsei.ac.kr}\orcid{0000-0002-1031-0796},
	Chul~Chung\orcid{0000-0001-6812-4542},
	Young-Wook~Lee\orcid{0000-0002-2210-1238},
	Seunghyun~Park\orcid{0009-0002-1351-1582},
	Junhyuk~Son\orcid{0009-0004-3117-1977},
	and Suk-Jin~Yoon\orcid{0000-0002-1842-4325}
}

\institute{
		Department of Astronomy \& Center for Galaxy Evolution Research, Yonsei University, Seoul 03722, Republic of Korea
}

\date{Received 31 December, 2025; accepted 09 September, 2026}

 \abstract {
Type Ia supernovae (SNe Ia) observables, such as the light-curve shape and the colour, are expected to contain information about the progenitor star.
In this work, we explore this information, with a particular focus on the age of the SN Ia progenitor star.
For this, we construct the SN Ia progenitor age distribution (SPAD) and compare it to the observed distributions of light-curve shape ($x_1$) and the colour ($c$) parameters in a volume-limited SN Ia sample.
We find that SPAD and the $x_1$ distribution share a common shape: a young/high-$x_{1}$ peak and an old/low-$x_1$ bump in the tail,  and this shape varies systematically with redshift.
In contrast, this behaviour is not evident in the $c$ distribution.
We then examine the correlation of the local age at the SN Ia explosion site, used as a proxy for the progenitor age, with $x_1$ and $c$.
The local age and $x_1$ are well correlated (the linear correlation coefficient $\simeq -0.71$), whereas the local age and $c$ show no significant correlation (the coefficient $\simeq 0.08$).
Furthermore, we find that the $x_1$ distribution systematically evolves with the local age.
Lastly, we demonstrate that an empirical mapping approach based on SPAD successfully reproduces the observed $x_1$ distribution across different redshift bins.
Taken together, our results suggest that the light-curve shape distribution indicates progenitor age at the population level more robustly than the colour does. 
In particular, younger progenitors are more likely to have higher-$x_1$ SNe Ia.
We discuss an application for creating a more homogeneous sample of SNe Ia in terms of progenitor age across a wide redshift range without the Malmquist bias, thereby improving the accuracy of cosmological constraints derived from SNe Ia.
}
 
\keywords{supernovae: general -- galaxy: evolution -- cosmology: observations -- distance scale
               }

\maketitle

\section{Introduction}
\label{sec:intro}

From an observation of a Type Ia supernova (SN Ia), we obtain light-curve data over time.
Then, the observed data are analyzed by light-curve fitters, which determine light-curve shape and colour parameters.
Through SN Ia standardization using these parameters, based on the empirical bright-slower and bright-bluer correlations \citep{Phillips1993, Tripp1998}, distance measurements can be achieved with an accuracy of $\sim$$7\%$.
Based on this accuracy and the assumption that the standardised luminosity is independent of the redshift and SN environments, the accelerating expansion of the Universe was discovered \citep{Riess1998, Perlmutter1999}.

However, as we enter the era of sub-percent precision cosmology \citep[e.g.,][]{Riess2022}, understanding the underlying physics of these empirical correlations has become critical for achieving unbiased cosmological parameter inference.
It is widely accepted that these empirical correlations are rooted in the physical properties of the SN Ia, such as the progenitor systems and their explosion mechanisms \citep[see][for recent reviews]{Liu2023, Ruiter2025}.
Because direct observations of SN Ia progenitor stars are difficult to make, even for the closest SN Ia (e.g., 2014J, see \citealt{Kelly2014}), previous studies employed the properties of host galaxies as a proxy for the properties of the progenitor stars.
They investigated the systematic correlations between light-curve parameters and SN environments, including the stellar mass \citep[e.g.,][]{Sullivan2010, Pan2014, Kim2019}, star formation rate \citep[e.g.,][]{Rigault2020, Kim2024a, Ramaiya2025}, rest-frame colours \citep[e.g.,][]{Roman2018, Kelsey2021, Kelsey2023}, cluster environments \citep[e.g.,][]{Toy2023, Aubert2025, Ruppin2025}, galactocentric distance \citep[e.g.,][]{Toy2025}, morphology from host galaxy image decomposition \citep[e.g.,][]{Senzel2025}, birth environments of the progenitor stars \citep[e.g.,][]{Kim2024b} and others.
In general, SNe Ia have a higher-stretch value when exploding in blue, star-forming, less massive ($\leq$ $10^{10}$ $M_{\odot}$), field galaxies and/or outer regions of galaxies, whereas the colour of SNe Ia has no or weak correlation.

Regarding the light-curve shape parameter, numerous host galaxy studies established a clear environmental dependency: SNe Ia that exploded in the less massive or star-forming or blue environments have a higher mean value of the SN light-curve shape parameter (i.e. intrinsically brighter) than those in massive or passive or red environments \citep[e.g.,][]{Sullivan2010, Pan2014, Kim2019, Kelsey2021, Senzel2025}. 
Considering the differences in stellar populations between those galaxy/environment properties, this result suggests that two populations exist for the SN Ia progenitor star.
This discussion of the two populations of SN Ia progenitor stars is also suggested in the SN Ia rate studies, as the prompt-and-delayed model \citep{Mannucci2005, Sullivan2006, Smith2012}.
They proposed that the observed relative rate of SN Ia in host galaxies could be explained by the existence of two populations, such that a younger SN progenitor population essentially depends on the star formation activity of host galaxies, and an older population depends on the host stellar mass.
Combining the above results, we can infer that a progenitor star of an SN Ia formed in a star-forming environment is more likely younger (on the order of tens/hundreds of Myr) with a higher light-curve shape value, and that formed in a passive environment might be older (Gyr-scale) with a lower light-curve shape value.
If so, since the galactic star formation rate evolves with redshift \citep{Madau2014}, the redshift evolution of the light-curve shape parameter is expected \citep{Howell2007}.
This was demonstrated at the 5$\sigma$ confidence level by \citet{Nicolas2021}, who studied an empirical description of the light-curve shape parameter (e.g., SALT2.4 $x_1$; \citealt{Guy2007, Guy2010, Betoule2014}) as a function of the redshift based on the volume-limited sample in $0.02 < z < 0.6$.
Recently, their empirical form of $x_1$ was confirmed by \citet{Ginolin2025a} based on an independent sample of 1,000 SNe Ia from the volume-limited ($z<0.06$) Zwicky Transient Facility \citep[ZTF;][]{Bellm2019, Graham2019} SN Ia DR2 \citep{Rigault2025}.

In contrast, regarding the SN Ia  colour parameter, no signs of significant differences in the mean colour values in different environments have been found \citep[e.g.,][]{Roman2018, Kim2019, Rigault2020, Kelsey2023, Ramaiya2025}.
However, we cannot simply conclude that this result is related to the different populations in the SN Ia progenitor star, as shown in the light-curve shape parameter, because the colour parameter is a mixture of intrinsic SN colour and the extrinsic effect, such as reddening by the dust of the host galaxy.
\citet{Ginolin2025b} attempted to separate the intrinsic SN colour from the extrinsic effect by modelling the observed colour, specifically SALT2.4 $c$, distribution as the convolution of a Gaussian intrinsic SN Ia colour and an exponential dust component based on the ZTF SN Ia DR2 sample.
From this, they showed that the intrinsic SN Ia colour is independent of the local and global environments.
This result is further confirmed by \citet{Popovic2025} with a larger sample of about 3,000 SNe Ia over a wider redshift range from 0.015 to 0.36, including the ZTF SN Ia DR2 sample.
They also found that the distribution of the intrinsic colour term does not change with redshift, while that of the extrinsic dust term varies with redshift at a confidence of >6$\sigma$.

To summarise, there is extensive observational evidence that the SN Ia light-curve shape parameter correlates with host environments and evolves with redshift, whereas the (intrinsic) colour does not.
This suggests that the light-curve shape parameter may contain more information about the SN Ia progenitor star.
This motivated \citet{Wojtak2023} and \citet{Wojtak2025} to develop the two-population Bayesian hierarchical model of SNe Ia; a similar two-population framework was independently proposed by \citet{Rubin2026}, called UNITY1.8.
As described in these studies, their two-population models were drawn \textit{probabilistically} from the distribution of stretch parameters.
Specifically, \citet{Wojtak2023} and \citet{Wojtak2025} assumed that SN local environments \citep[e.g.,][]{Rigault2020} are differentiated based on population type with respect to the stretch parameter, and referred to the populations associated with the high- and low-stretch peaks of the stretch distribution as young and old populations.
As they noted, this association did not come from a direct comparison with the progenitor age.

We expect that such a comparison provides a direct, quantitative test of whether the observed light-curve shape and colour distributions, and their redshift evolutions, are consistent with the progenitor star age distribution.
It would also establish which SN Ia observables best indicate the progenitor age.
Despite this need, no study has directly performed this comparison, because it is highly challenging to directly observe the progenitor star, and thus its properties cannot be determined observationally\footnote{See \citet{Kim2024b} for an indirect approach to estimating the properties of the progenitor star. They introduced a method to trace back the birth environments, such as metallicity, of the SN Ia progenitor stars. Considering the birth environments as the properties of the progenitor stars, \citet{Kim2025} attempted to constrain the progenitor scenarios of SNe Ia.}.
However, a method for deriving the SN Ia progenitor star age distribution as a function of redshift was introduced by \citet{Childress2014}, who convolved empirical models of galaxy mass assembly histories with theoretical delay time distribution models for SNe Ia (see also \citealt{Wiseman2022}, for further developments of this method).

Therefore, in the present work, first we construct the SN Ia progenitor star age distribution as a function of redshift based on \citet{Childress2014}'s method and directly compare its predicted age structure with the observed light-curve shape and colour distributions.
Second, we test this interpretation empirically using local stellar population ages at SN Ia explosion sites, assuming that these local ages trace the SN Ia progenitor star ages.
Lastly, we employ an empirical mapping approach to evaluate whether the observed light-curve shape distribution and its redshift evolution can be reproduced from the SN Ia progenitor star age distribution.
Together, these analyses allow us to assess which SN Ia observable indicates progenitor age most robustly at the population level.

The remainder of this paper is structured as follows.
In Section~\ref{sec:method}, we describe the method for constructing the progenitor star age distribution and present the observational sample and empirical descriptions of SN Ia observables.
In Section~\ref{sec:results}, we present the comparison between the progenitor star age distribution and observed light-curve shape and colour parameters, including comparisons using local stellar population ages around the SN Ia explosion site.
In Section~\ref{sec:forward_model}, we predict the observed light-curve shape distribution through an empirical mapping approach based on  the progenitor star age distribution.
Applications of these results and directions for future works are discussed in Section~\ref{sec:discussion}.

\section{Methodology}
\label{sec:method}

\subsection{Constructing the SN Ia progenitor star age distribution}
\label{subsec:method_spad}

\begin{figure}
  \centering
  \includegraphics[width=\columnwidth]{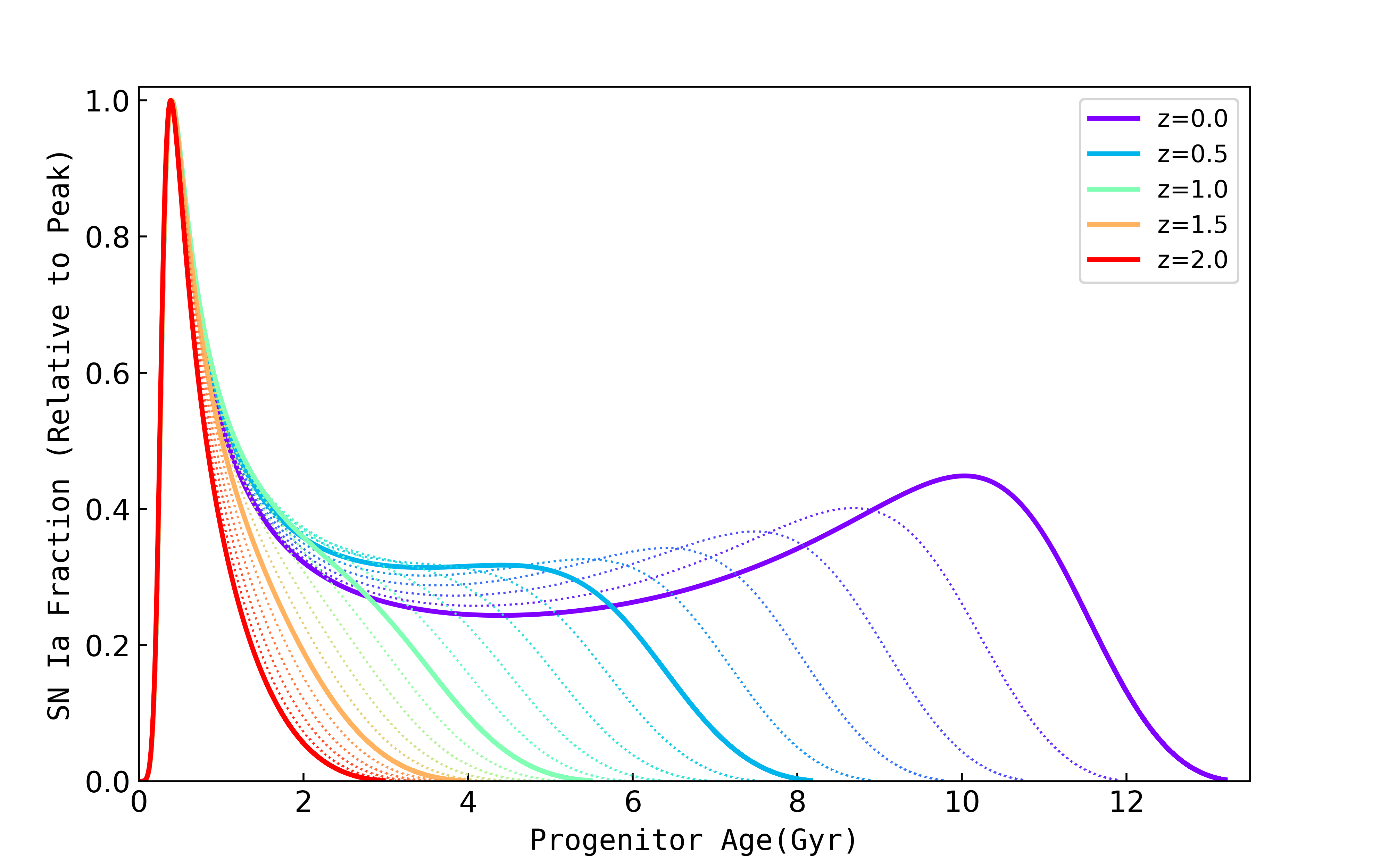}
  \caption{Constructed SN Ia progenitor age distributions as a function of redshift from $z = 0.0$ (purple) to $z = 2.0$ (red) in a step of $\Delta z = 0.1$.
  	Following the \citetalias{Childress2014} method, the SN Ia progenitor age distribution is derived by the convolution of the SN Ia DTD model (Eq.~\ref{eq.dtd}) with the cosmic SFH (Eq.~\ref{eq.csfr}).
	}
\label{fig:spad_z}
\end{figure}

In order to construct the average SN Ia progenitor star age distribution (hereafter SPAD) at a given redshift, we follow the method as described in \citet[hereafter \citetalias{Childress2014}]{Childress2014}, which is the method adopted in several previous studies \citep[e.g.,][]{Kang2020, Lee2022, Son2025}.
The SPAD at epoch $t_{0}$ for the SN arising from a progenitor system of age $\tau$, $P(\tau; t_{0})$, is derived as the integrand of  the convolution of the SN Ia delay time distribution (DTD; $\phi(\tau)$) model with the galaxy star formation history (SFH; $\psi(t_{0} - \tau)$) , in the form of 
\begin{equation} \label{eq.spad}
P(\tau; t_{0}) = \phi(\tau)\psi(t_{0} - \tau).
\end{equation}

For the SN Ia DTD model, we employed a smooth functional form as suggested by \citetalias{Childress2014}:
\begin{equation} \label{eq.dtd}
\phi(t) \propto \frac{(t/t_{\text{prompt}}) ^{\alpha}}{(t/t_{\text{prompt}}) ^{\alpha-s} + 1},
\end{equation}
where $t_{\text{prompt}}$ is the characteristic `prompt' population timescale and $s$ is the power law slope with a high-order polynomial of $\alpha$.
At late times ($t \ge t_{\text{prompt}}$), this form rapidly approaches $t^s$ power law, while at early times, it is a high-order polynomial which flattens to zero rapidly below $t_{\text{prompt}}$. 
We use $s=-1$, which is supported by numerous observational studies across a wide range of environments and redshifts \citep[e.g.,][]{Maoz2010, Maoz2012, Graur2011, Graur2014, Frohmaier2019}.

Different binary population synthesis models predict a broad prompt timescale ranging from $\sim$40 Myr to a few hundred Myr (see figure 5 of \citealt{Wang2012} and references therein).
\citetalias{Childress2014} examined different SPADs with these various forms of the SN Ia DTD and found that their qualitative behaviour follows their baseline SPAD with  $t_{\text{prompt}} = 0.3$ Gyr.
The combination of $t_{\text{prompt}}= 0.3$ Gyr and the high-order polynomial of $\alpha = 5$ we adopted naturally reproduces this broad prompt peak while smoothly suppressing the rate at extreme early times, providing a more realistic representation of binary evolution.

For the SFH of the galaxy, instead of determining individual SFH of each galaxy, we take an ensemble average of SFHs of all types of galaxies at a given redshift (i.e., cosmic time) to calculate the average SPAD at a given redshift.
The empirical approach of \citet{Madau2017} from the collected cosmic SFR density (CSFR) is employed (see their eq.(1)),
\begin{equation} \label{eq.csfr}
\text{CSFR}(z) = 0.01\frac{(1+z)^{2.6}}{1 + [(1+z) / 3.2]^{6.2}} M_{\odot} \text{ yr}^{-1} \text{ Mpc}^{-3}.
\end{equation}
They described that this formalism is an updated version of the formula given in \citet{Madau2014} that better reproduces a number of $4 \le z \le 10$ results.

Fig.~\ref{fig:spad_z} presents the constructed SPAD as a function of redshift derived from multiplying the SN Ia DTD model (Eq.~\ref{eq.dtd}) by the mean SFHs at a given redshift (Eq.~\ref{eq.csfr}), successfully reproducing \citetalias{Childress2014}.
As pointed out by \citetalias{Childress2014}, we can also observe the apparent bimodal distribution of SPAD from $z=0.0$ to $z=0.5$, which indicates the ``prompt'' (or young) and ``delayed'' (or old) populations of SNe Ia.
From $z>0.5$, SPAD begins having only one peak at $\sim$0.4 Gyr which is a similar value to $t_{\text{prompt}}$ (= 0.3 Gyr) for the DTD model.
This would imply that there is only one ``prompt'' population in the higher redshift range.
This can be naturally explained by the fact that as the redshift increases, the age of the Universe and thus the average age of stellar populations become younger. 
Consequently, higher redshift SNe Ia can only originate from young progenitor stars, whereas both young and old stars at lower redshifts contribute to the SN Ia populations.

We note that adopting a shorter $t_{\text{prompt}}$ (e.g., 0.05 or 0.1 Gyr) slightly decreases the relative amplitude of the old population, as expected from the larger fraction of the young population.
However, it does not materially affect our main results, as the overall bimodal shape of SPAD and its systematic redshift evolution remain robust.

\subsection{Sample}
\label{subsec:sample}

\begin{table}
\centering
\caption{The summary of our sample with a redshift cut ($z_{\text{lim}}$) for the volume-limited sample.
		Note that the HST sample is not volume-limited (see Sec.~\ref{subsec:sample}).}
\label{tab:sample}
\begin{tabular}{l c c c }
\hline\hline\\[-0.8em]
\multicolumn{2}{c}{Survey} & $z_{\text{lim}}$ & $N_{\text{SNIa}}$  \\[0.15em] 
\hline\\[-0.8em]
\multicolumn{2}{c}{ZTF SN Ia DR2} 			& 0.06 & 890  \\[0.30em]
\hline\\[-0.8em]
\multirow{5}{*}{\citetalias{Nicolas2021} Dataset}  	&SNf & 0.08 & 111  \\[0.30em]
								& SDSS & 0.20 & 167  \\[0.30em]
								& PS1 & 0.31 & 160   \\[0.30em]
								& SNLS & 0.60 & 102  \\[0.30em]
								& HST & - & 26  \\[0.30em]
\hline
\multicolumn{2}{c}{Total} & & 1456  \\
\hline
\end{tabular}
\end{table}

For the SN Ia observables, we employ SALT2.4 light-curve stretch ($x_1$) and colour ($c$) parameters.
We took these parameters from a volume-limited sample of ZTF SN Ia DR2 and an SN Ia dataset of \citet[hereafter \citetalias{Nicolas2021}]{Nicolas2021}.
Tab.~\ref{tab:sample} summarises our sample.
We note that the ZTF SN Ia DR2 sample used SALT2.4 retrained by \citet{Taylor2021}.
However, the impact of a new surface on the light-curve parameters is small, because the effect is largely compensated for by the change in absolute magnitude, as discussed in \citet{Taylor2021}.

The volume-limited ZTF SN Ia DR2 sample is presented in \citet{Rigault2025}.
The initial DR2 sample contains 3628 spectroscopically confirmed nearby ($z<0.3$) SNe Ia.
After basic cuts they defined, 2667 SNe Ia remained.
To obtain a volume-limited sample, they limited their sample to SNe Ia at $z_{\text{lim}}<0.06$ based on the prescription from survey simulations \citep{Amenouche2025}.
This left 993 SNe Ia.
For the present work, we further limit the sample to $c < 0.3$, a cut for selecting cosmologically normal SNe Ia, while they used $c < 0.8$ for studying the $c$ distribution and its evolution \citep[e.g., ][]{Ginolin2025b, Popovic2025}.
The final number of the ZTF SN Ia DR2 sample we used is 890 SNe Ia.

As shown in Table~\ref{tab:sample}, \citetalias{Nicolas2021} dataset combines an SN Ia sample from the SDSS-II Supernova survey \citep[hereafter SDSS;][]{Smith2012, Sako2018}, the Pan-STARRS1 \citep[hereafter PS1;][]{Rest2014}, and the Supernova Legacy Survey \citep[hereafter SNLS;][]{Guy2010, Sullivan2010} collected in the Pantheon catalog \citep{Scolnic2018}.
Considering the selection effect in each survey carefully, they applied a redshift cut to make a volume-limited sample: for the SDSS sample $z_{\text{lim}}=0.20$ , for the PS1 sample $z_{\text{lim}}=0.31$, and for the SNLS sample $z_{\text{lim}}=0.60$.
For the HST sample, they do not impose further cuts because the target classification was robust enough to include them in the cosmological analysis, as investigated by \citet{Scolnic2018}\footnote{We note that the HST sample is therefore not volume-limited. Following \citetalias{Nicolas2021}, we retain it for a comparison at the highest redshift bin, where the SPAD prediction is most distinctive, and our main conclusion is unchanged when this sample is excluded. }.
Furthermore, they added the Nearby Supernova Factory \citep[hereafter SNf;][]{Aldering2020, Rigault2020} sample, which is volume-limited, to further cover SNe Ia at $z<0.1$.
In total, we used 566 SNe Ia in the \citetalias{Nicolas2021} dataset.

From this sample of 1456 SNe Ia (volume-limited except for the 26 HST data), we expect that we can observe a true trend.

\subsection{Empirical descriptions for the light-curve shape and the colour}
\label{subsec:x1_c}

\citetalias{Nicolas2021} provided the empirical redshift drift model of the underlying $x_1$ distribution of SNe Ia as a function of redshift ($x_1(z)$) from their dataset.
They explained that the $x_1$ probability distribution function of a given SN Ia will be the linear combination of the $x_1$ distributions of young and old populations, based on the local star formation rate determined by \citet{Rigault2020}, weighted by its probability to be young.
In general, the fraction of young SNe Ia (inferred from a locally star-forming environment) as a function of redshift is given by $\delta(z)$, which was introduced by \citet{Rigault2020}, such that

\begin{equation}
\label{eq:r20_agedrift}
\delta(z) = \left( K^{-1} \times (1 + z)^{-2.8} + 1 \right)^{-1}
\end{equation}

with $K$ = 0.87.
Therefore, \citetalias{Nicolas2021} suggested that their $x_1(z)$ is given by

\begin{multline}
\label{eq:n21_x1}
x_1(z) = \delta(z) \times \mathcal{N}(\mu_1, \sigma_{1}^{2}) \\ 
+ (1 - \delta(z)) \times \left[ a \times \mathcal{N}(\mu_1, \sigma_{1}^{2}) + (1 - a) \times \mathcal{N}(\mu_2, \sigma_{2}^{2}) \right]
\end{multline}

where $\mathcal{N}(\mu, \sigma)$ is the normal distribution, subscripts $1$ and $2$ mean the high-$x_1$ and low-$x_1$ mode, and $a$ is the relative effect of two modes.
They fit the model with their dataset and obtained $a = 0.51$, $\mu_1 = 0.37$, $\mu_2 = -1.22$, $\sigma_1 = 0.61$, and $\sigma_2 = 0.56$.
We refer the reader to \citetalias{Nicolas2021} for more detailed discussions of the derivation of the equation.

We employed the empirical SN Ia colour model given by \citet{Ginolin2025b}.
Following former analyses \citep[e.g.,][]{Jha2007, Mandel2017}, \citet{Ginolin2025b} provided the functional form of the colour model that is the convolution of a Gaussian intrinsic SN Ia colour ($\mathcal{N}(c | c_{int}, \sigma_c)$) with an exponential dust component ($\tau$), 

\begin{equation}
\label{eq:colour_model}
P(c) = \mathcal{N}(c | c_{int}, \sigma_c) \otimes \left\{ \begin{array}{lcl} 	0 & \mbox{if} & c \leq 0 \\
													 	\frac{1}{\tau}e^{-c/\tau} & \mbox{if} & c > 0.
										\end{array} \right.
\end{equation}

From the volume-limited sample of ZTF SNIa DR2 ($z<0.06$) with an extended colour range ($c < 0.8$), \citet{Ginolin2025b} determined $c_{int}$ = $-0.085\pm0.004$, $\sigma_c$ = $0.030\pm0.005$, and $\tau$ = $0.155\pm0.007$.
They mentioned that conceptually, the normal distribution should represent the intrinsic SN Ia colour scatter, while the exponential decay accounts for additional reddening due to the interstellar dust of the host galaxy.
For the sample at $z>0.1$, we adopt $c_{int}$ = $-0.085\pm0.004$ and $\sigma_c$ = $0.030\pm0.005$ as in \citet{Ginolin2025b} and $\tau = 0.110$ taken from \citet[their figure 4]{Popovic2025}.
This is because, as we discussed in Sec.~\ref{sec:intro}, they found that the distribution of the intrinsic colour term does not change with redshift, while that of the extrinsic dust term varies with redshift at a confidence of >6$\sigma$, with a larger sample of about 3,000 SNe Ia over a wider redshift range from 0.015 to 0.36.
As shown in their figure 4, the $\tau$ value (i.e., $E_{Dust}$) decreases from $\sim$0.17 at $z < 0.05$ to $\sim$0.11 at $z>0.1$, with the last four bins at $z=0.11$--$0.32$ being mutually consistent with a constant value.
We therefore adopt $\tau = 0.110$ for all $z>0.1$, assuming no further evolution beyond $z=0.36$.
We refer the reader to \citet{Ginolin2025b} and \citet{Popovic2025} for more detailed discussions of the derivation of the equation.



\begin{figure*}
  	\includegraphics[width=0.3\textwidth]{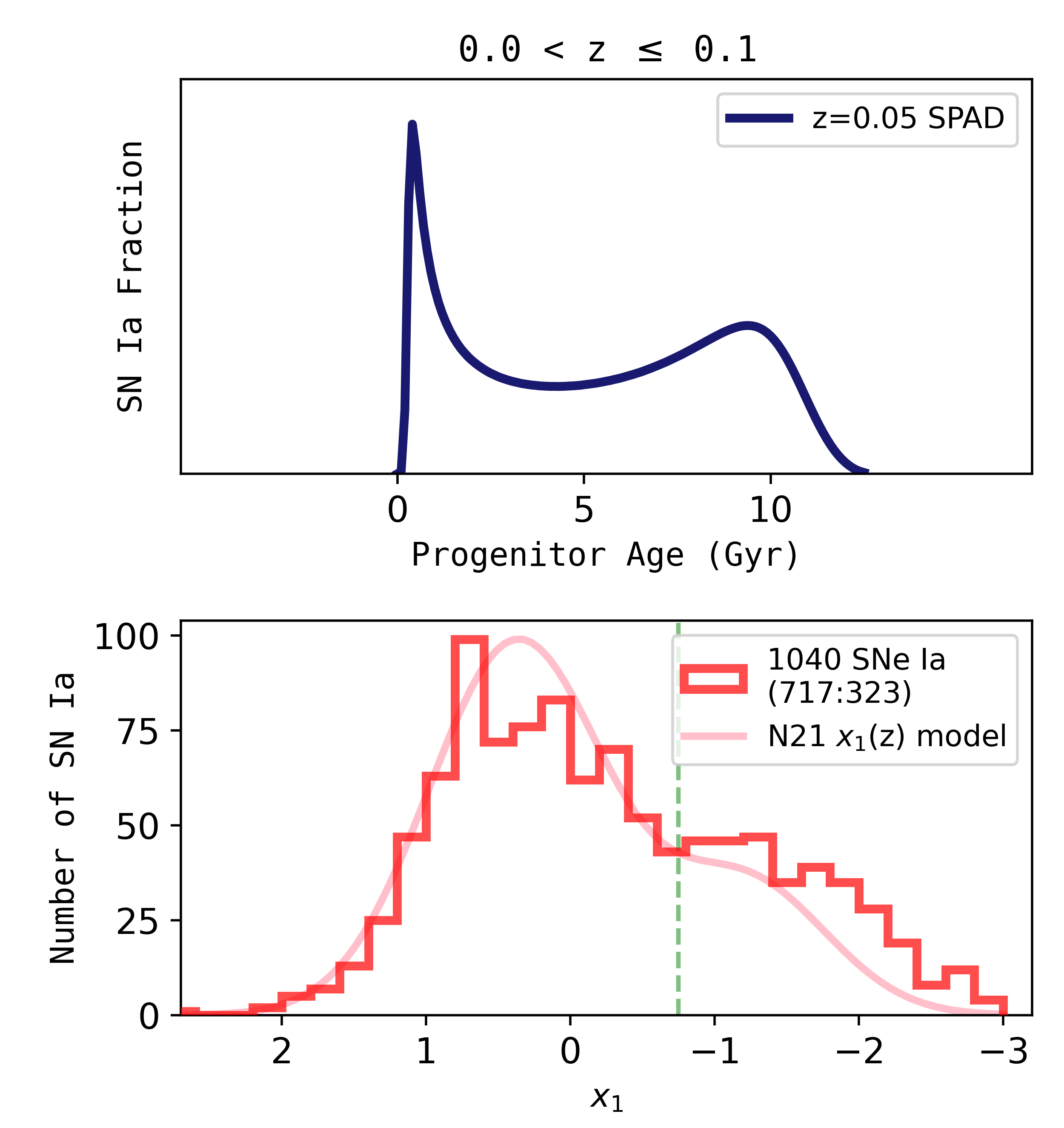}
  	\includegraphics[width=0.3\textwidth]{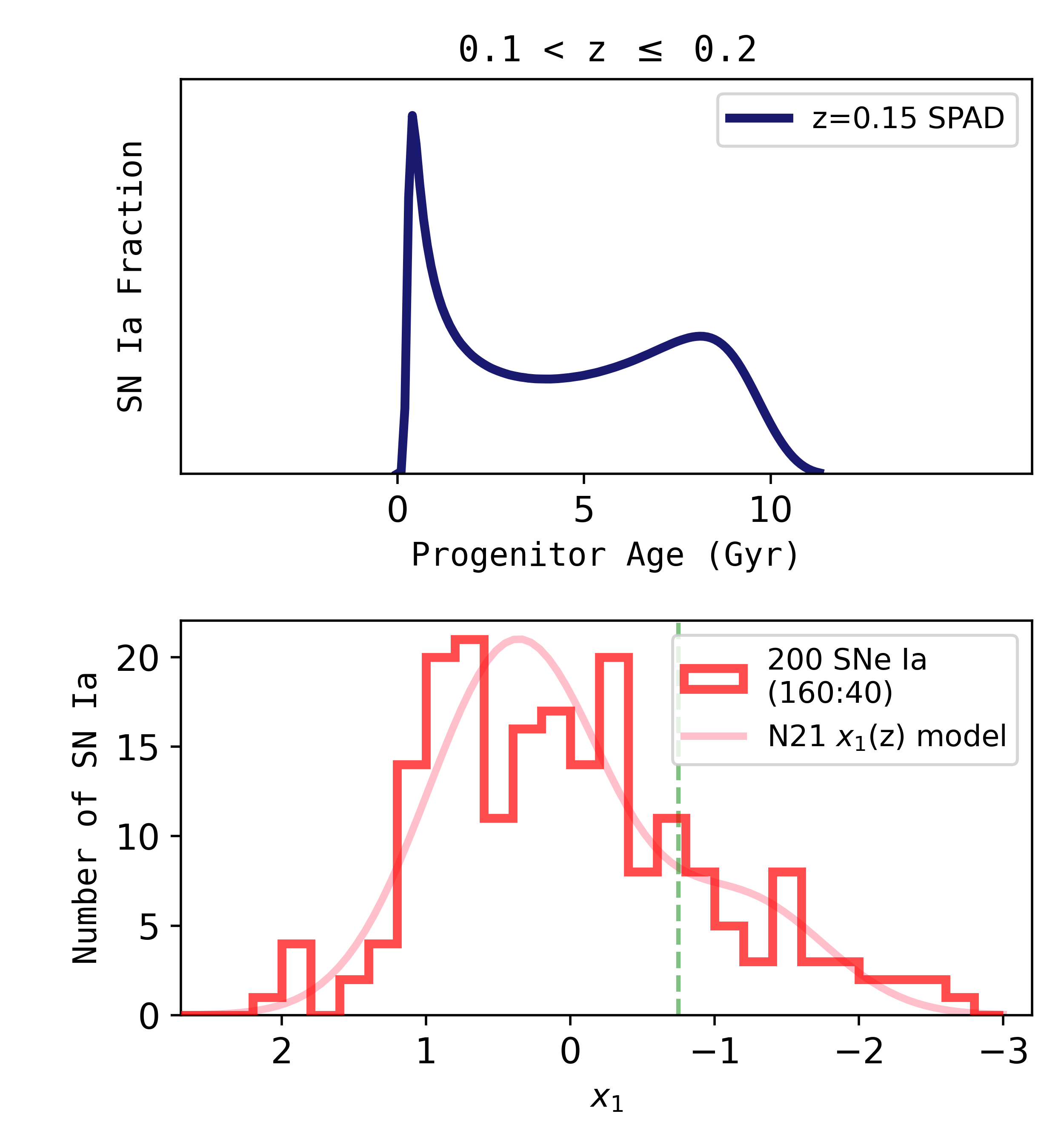}
  	\includegraphics[width=0.3\textwidth]{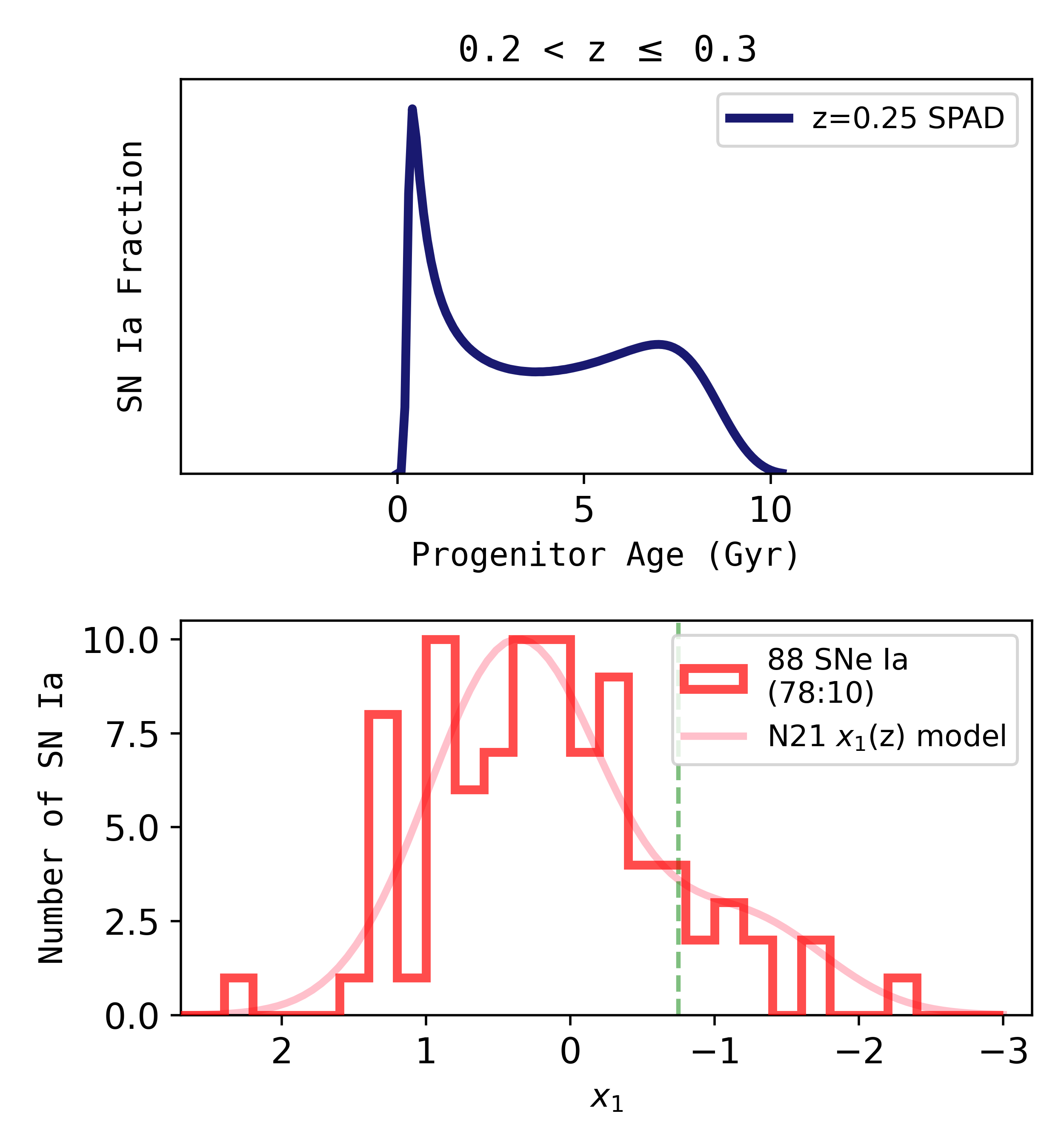}
  	\includegraphics[width=0.3\textwidth]{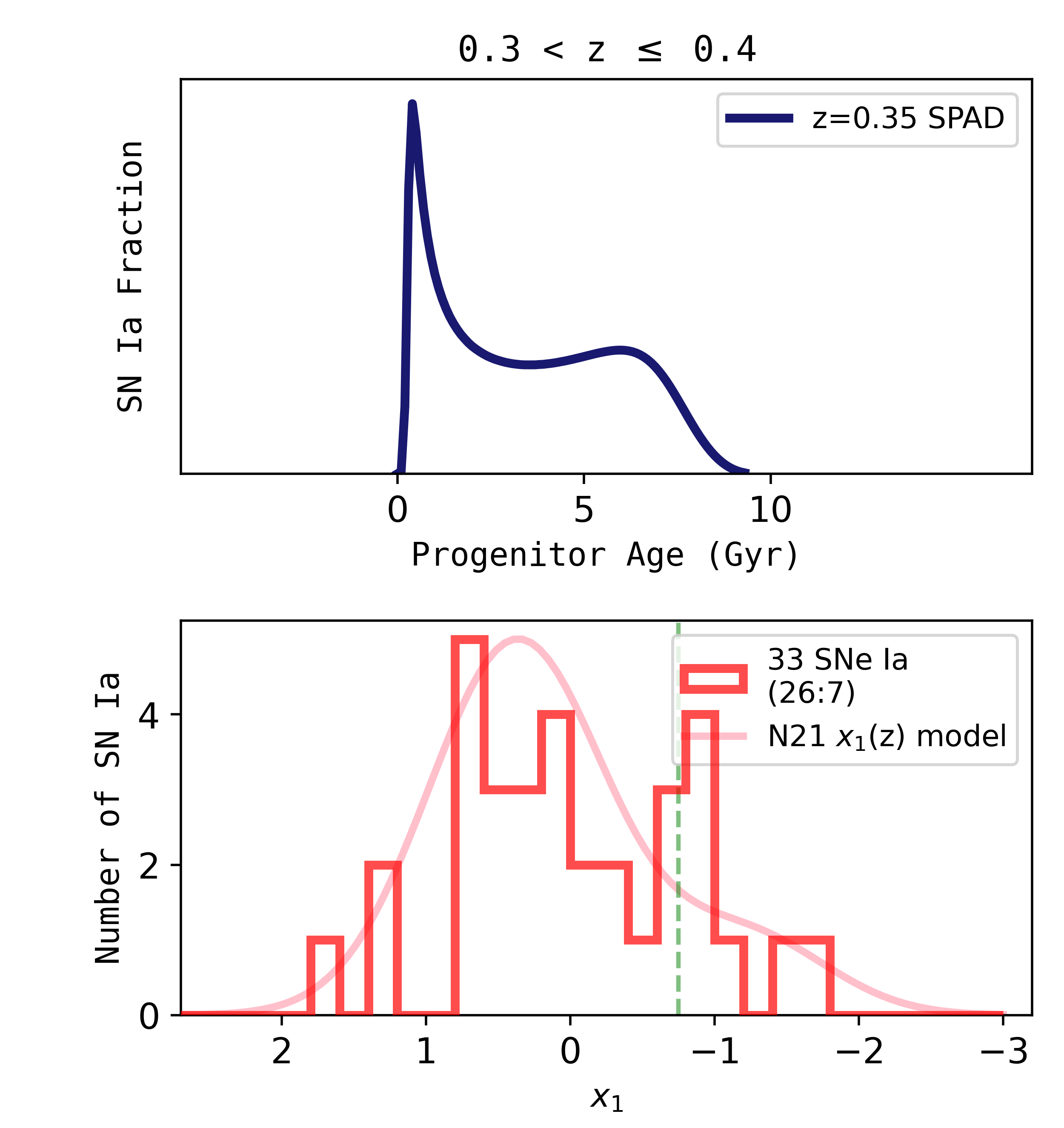}
  	\includegraphics[width=0.3\textwidth]{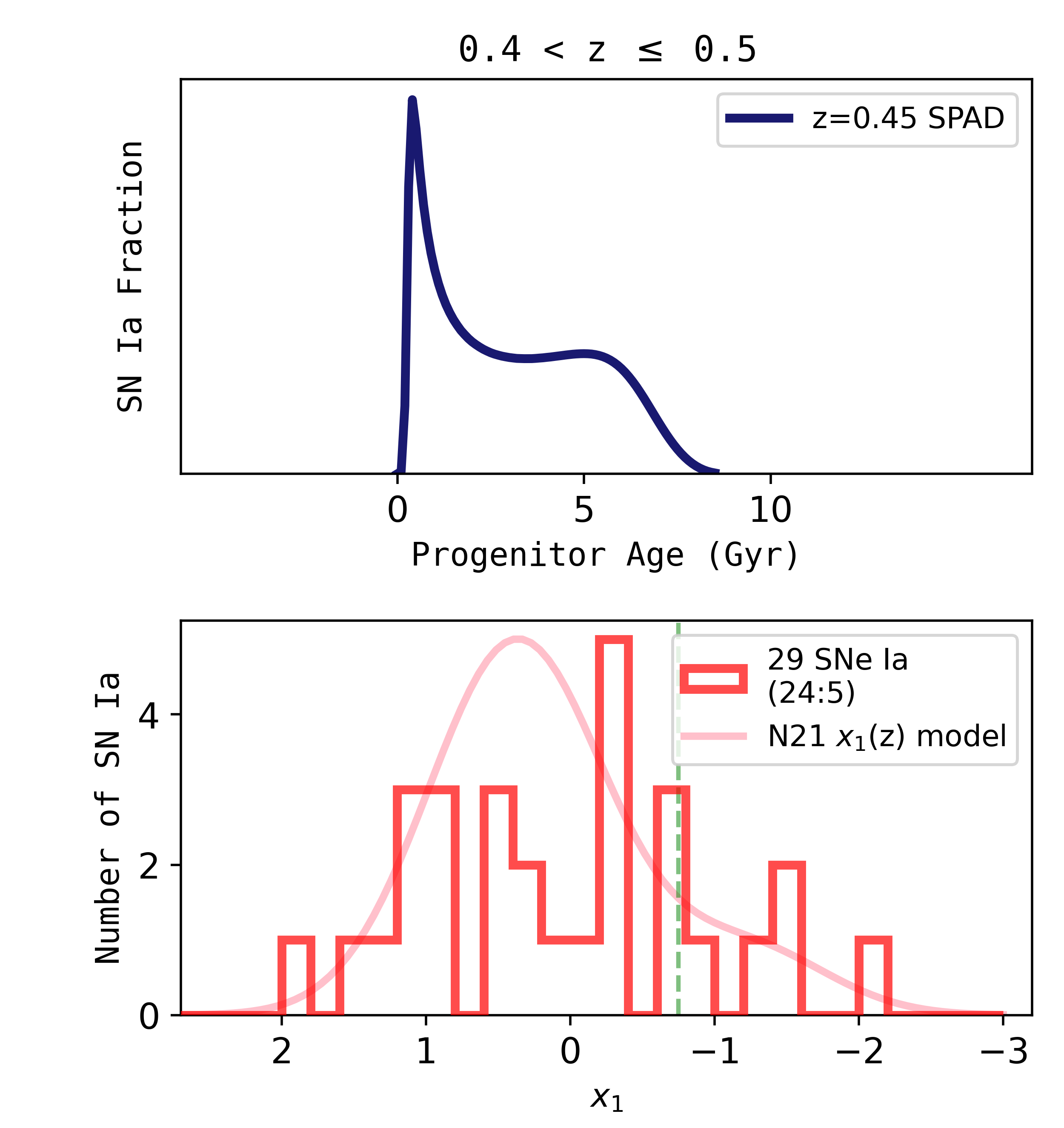}
  	\includegraphics[width=0.3\textwidth]{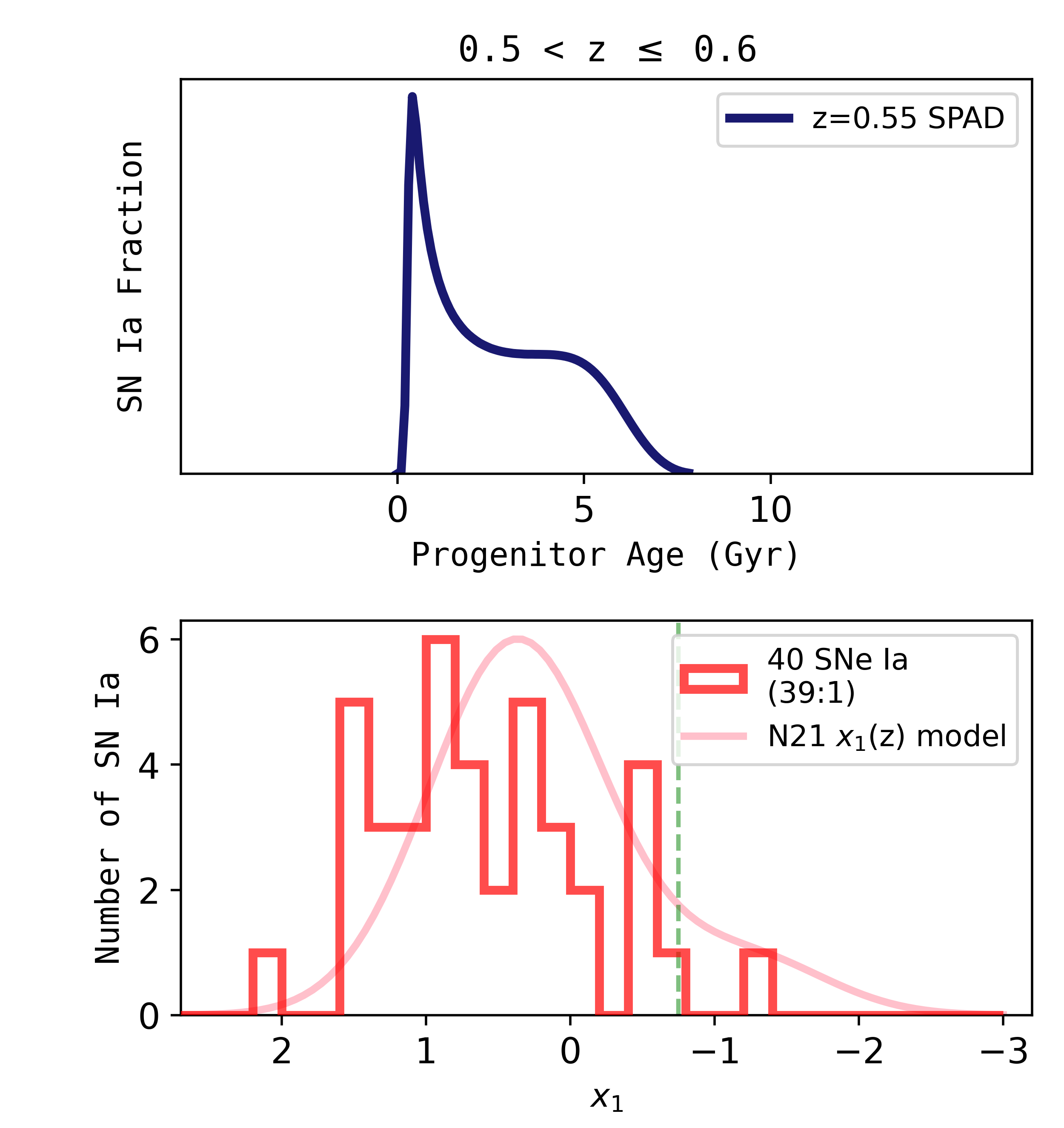}
  	\includegraphics[width=0.3\textwidth]{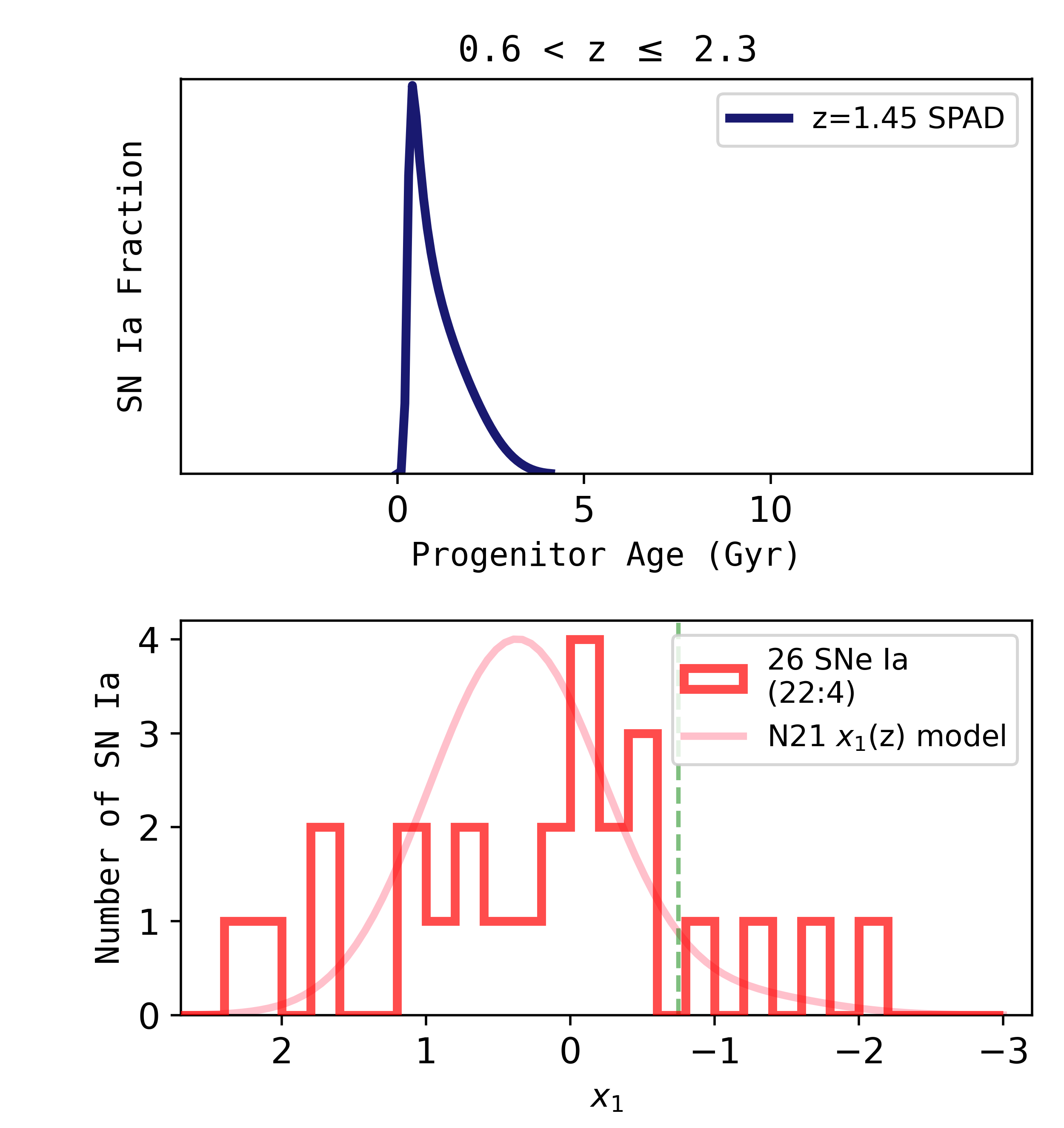}
  \caption{Distributions of observed SALT2.4 $x_{1}$ (bottom panel) and constructed SPAD (top) in different redshift bins.
  		The empirical \citetalias{Nicolas2021} $x_1$($z$) model (Eq.~\ref{eq:n21_x1}) is overplotted on the $x_1$ distribution with a pink solid line. 
		SPAD and the $x_1$($z$) model are derived with the middle value of each redshift bin.
		The green dashed vertical line indicates the split point at $x_1 = -0.75$, where the sample is divided into high- and low-$x_1$ SNe Ia.
		The number of the sample in each redshift bin and that of high- and low-$x_1$ SNe Ia are presented in the legend for each panel.
		We note that the $x_{1}$ value is flipped for comparison with SPAD.}
  \label{fig:x1_spad}
\end{figure*}

\begin{figure*}
  	\includegraphics[width=0.3\textwidth]{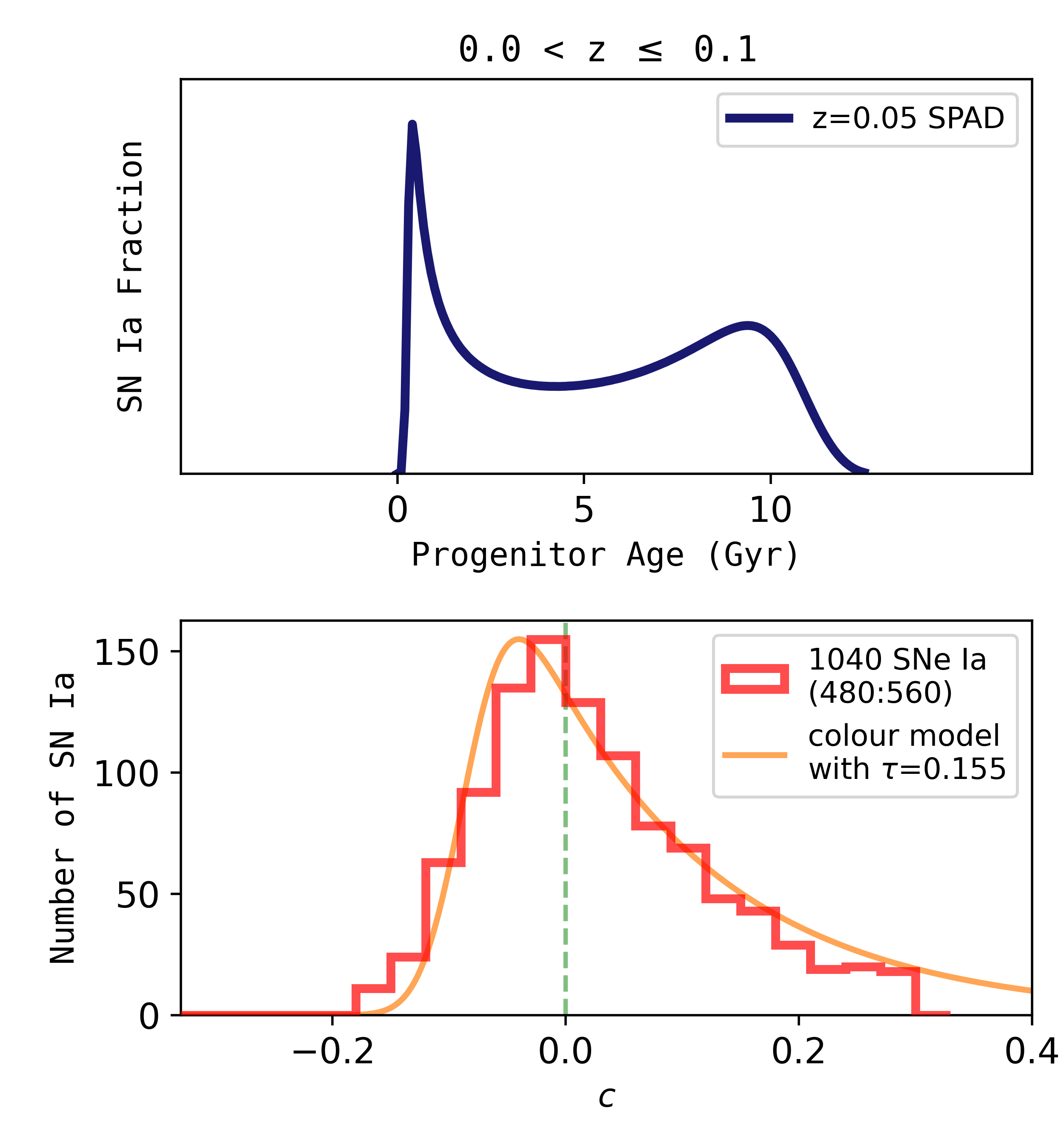}
  	\includegraphics[width=0.3\textwidth]{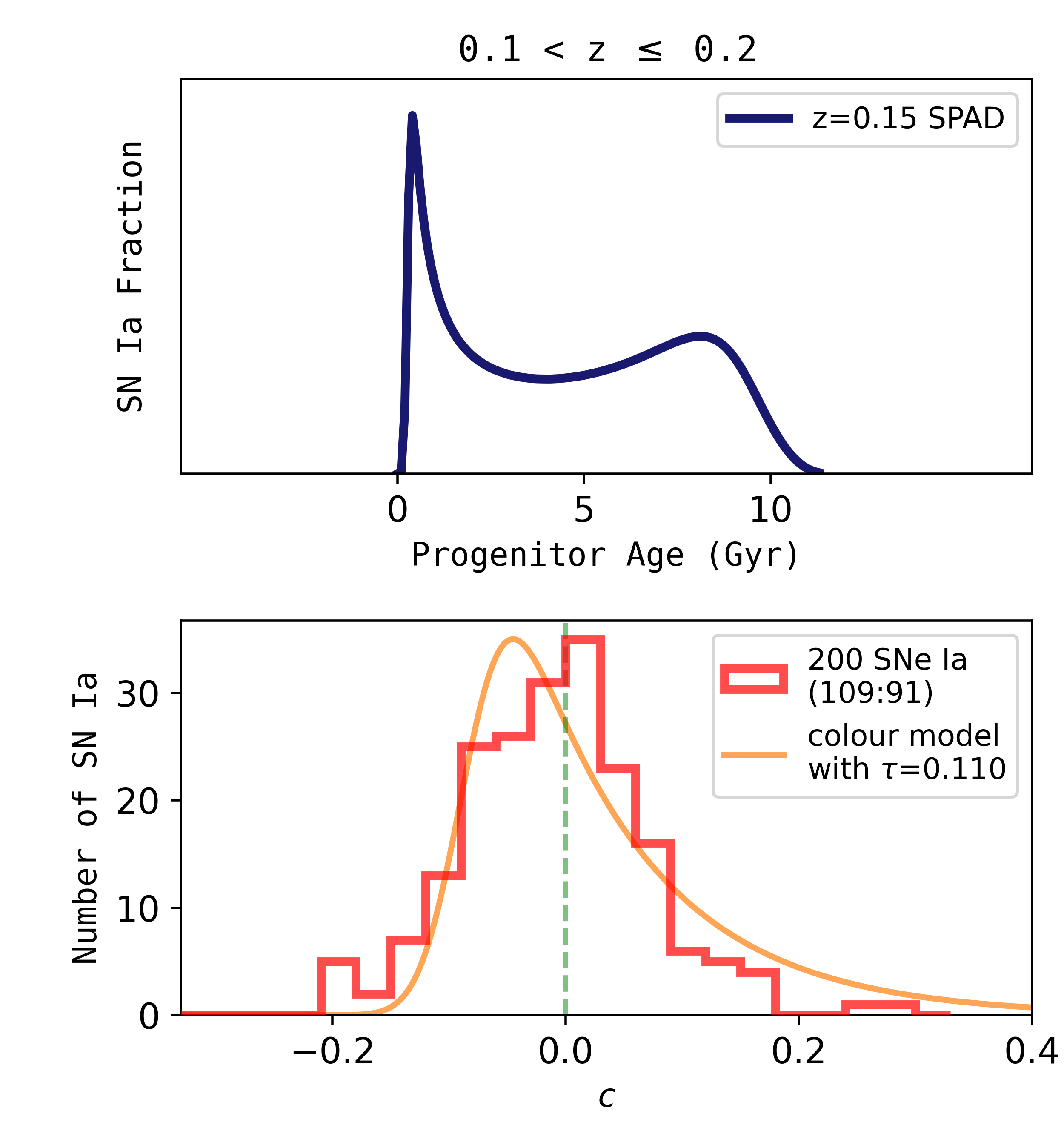}
  	\includegraphics[width=0.3\textwidth]{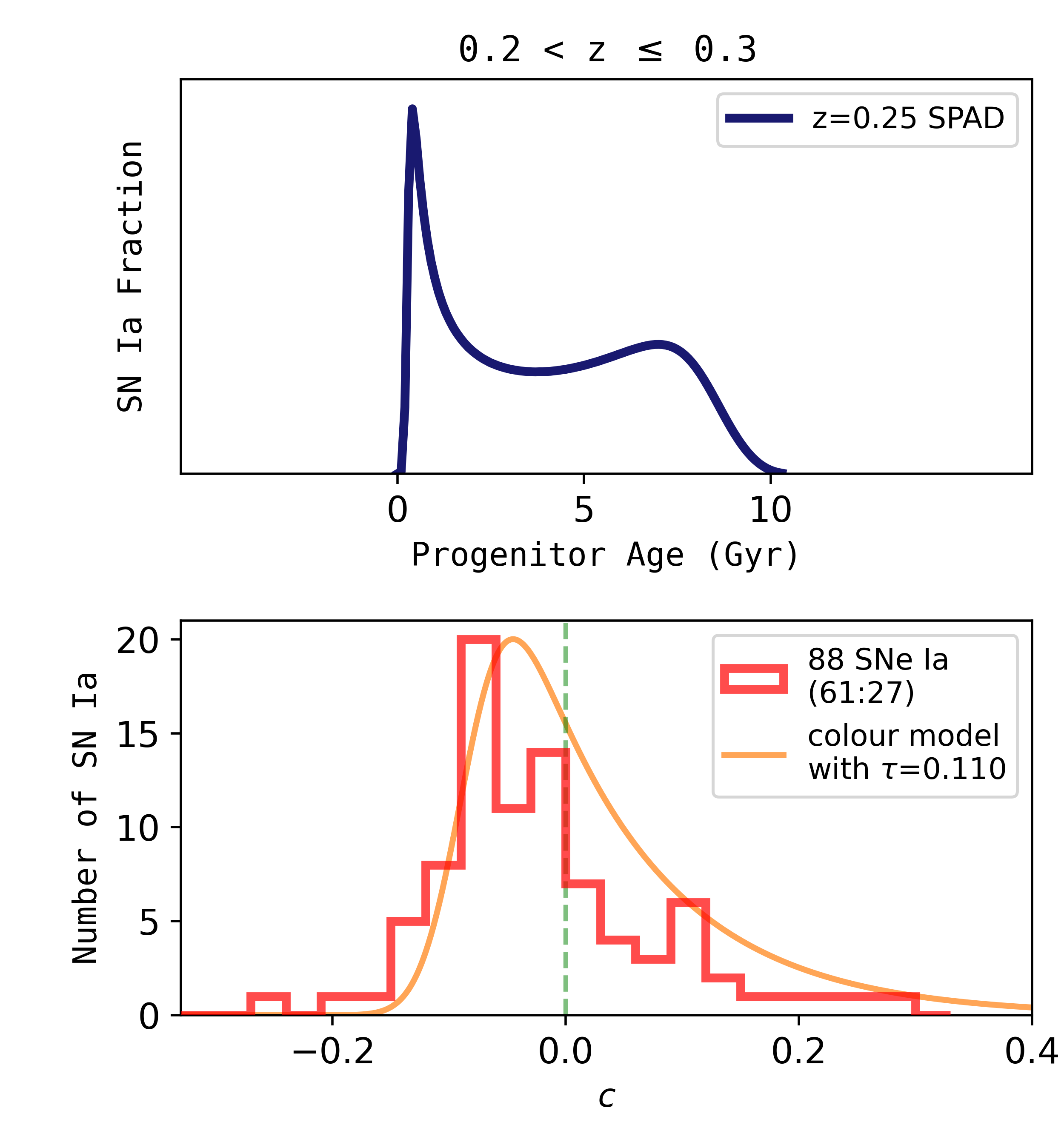}
  	\includegraphics[width=0.3\textwidth]{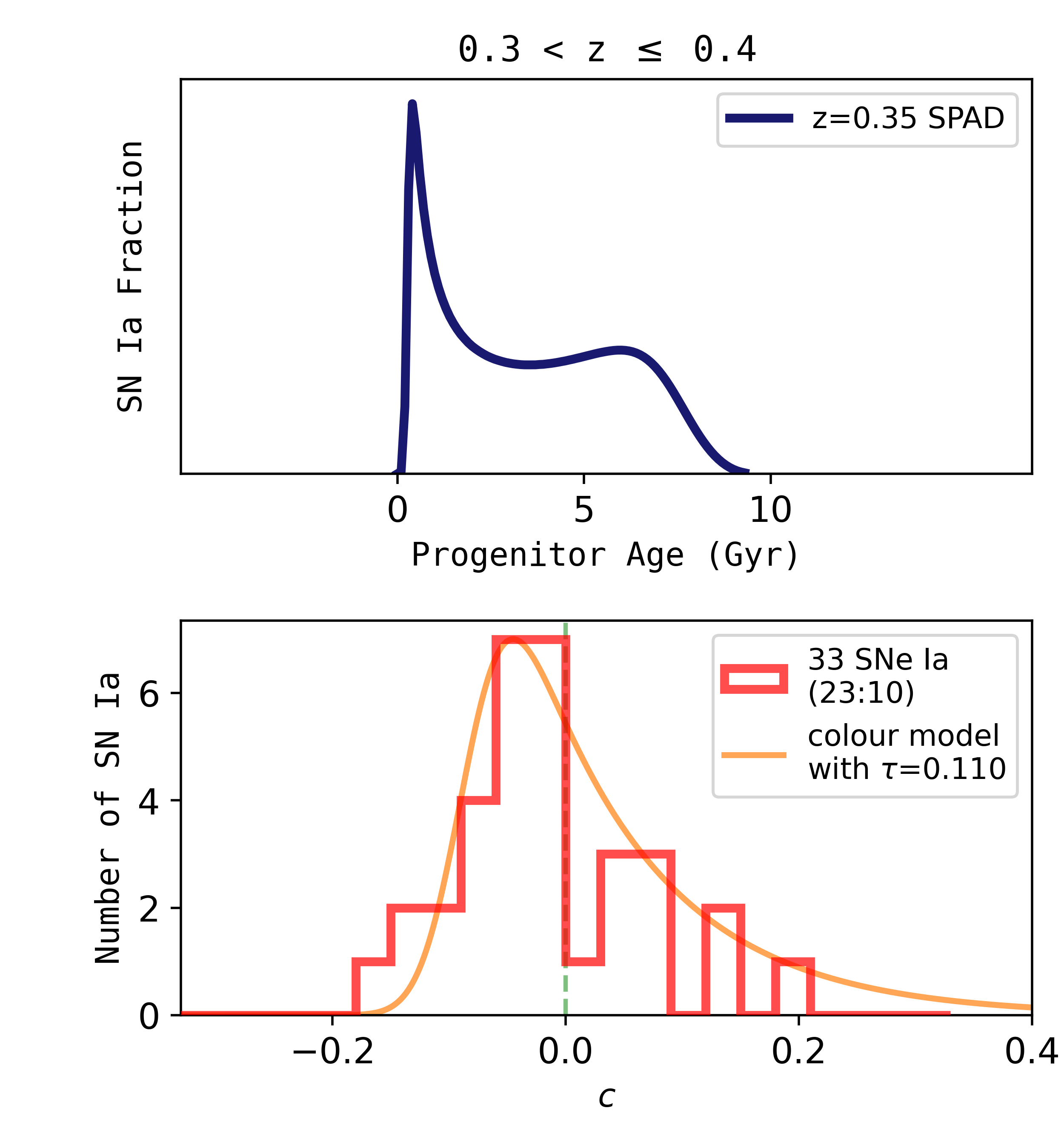}
  	\includegraphics[width=0.3\textwidth]{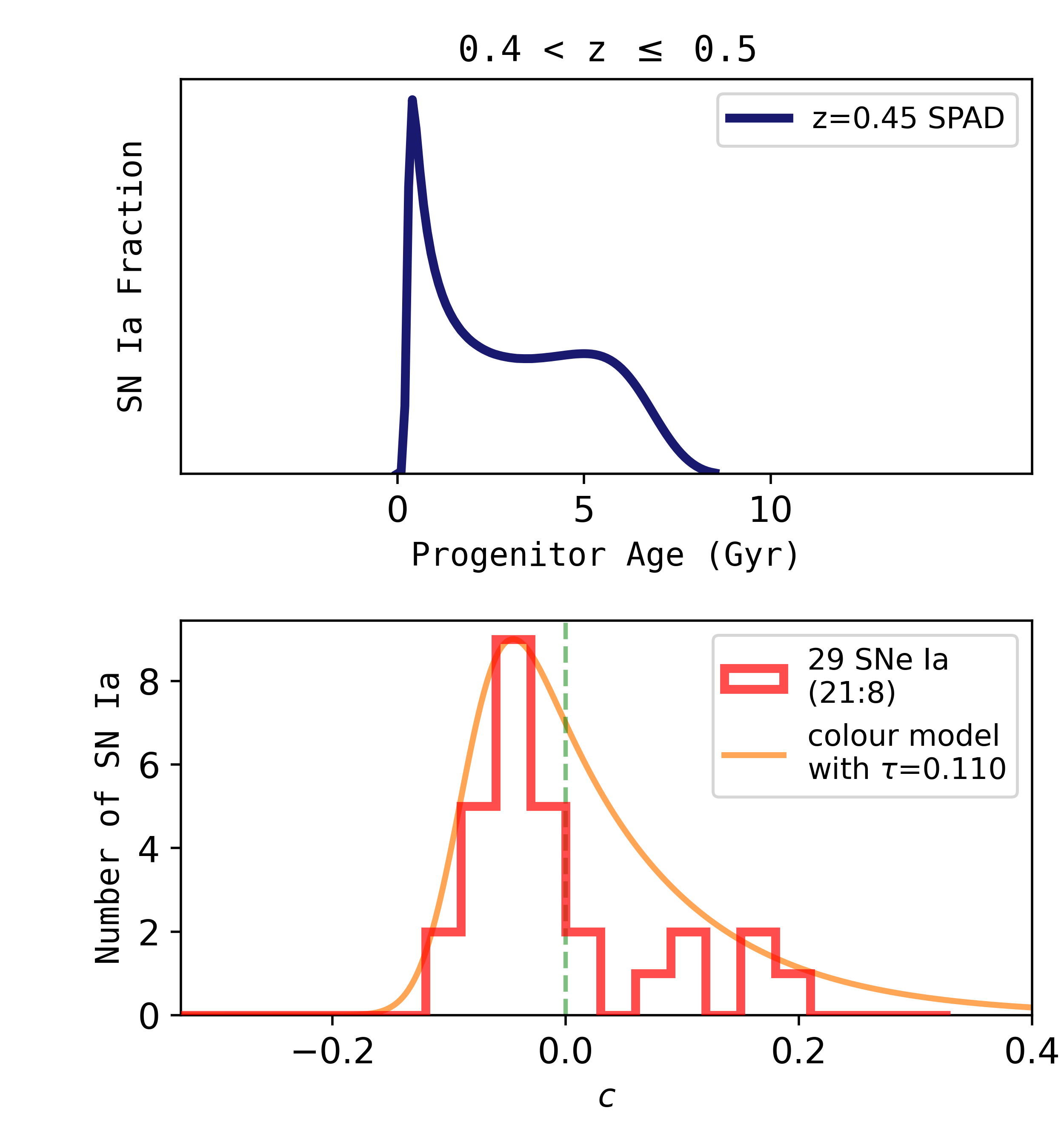}
  	\includegraphics[width=0.3\textwidth]{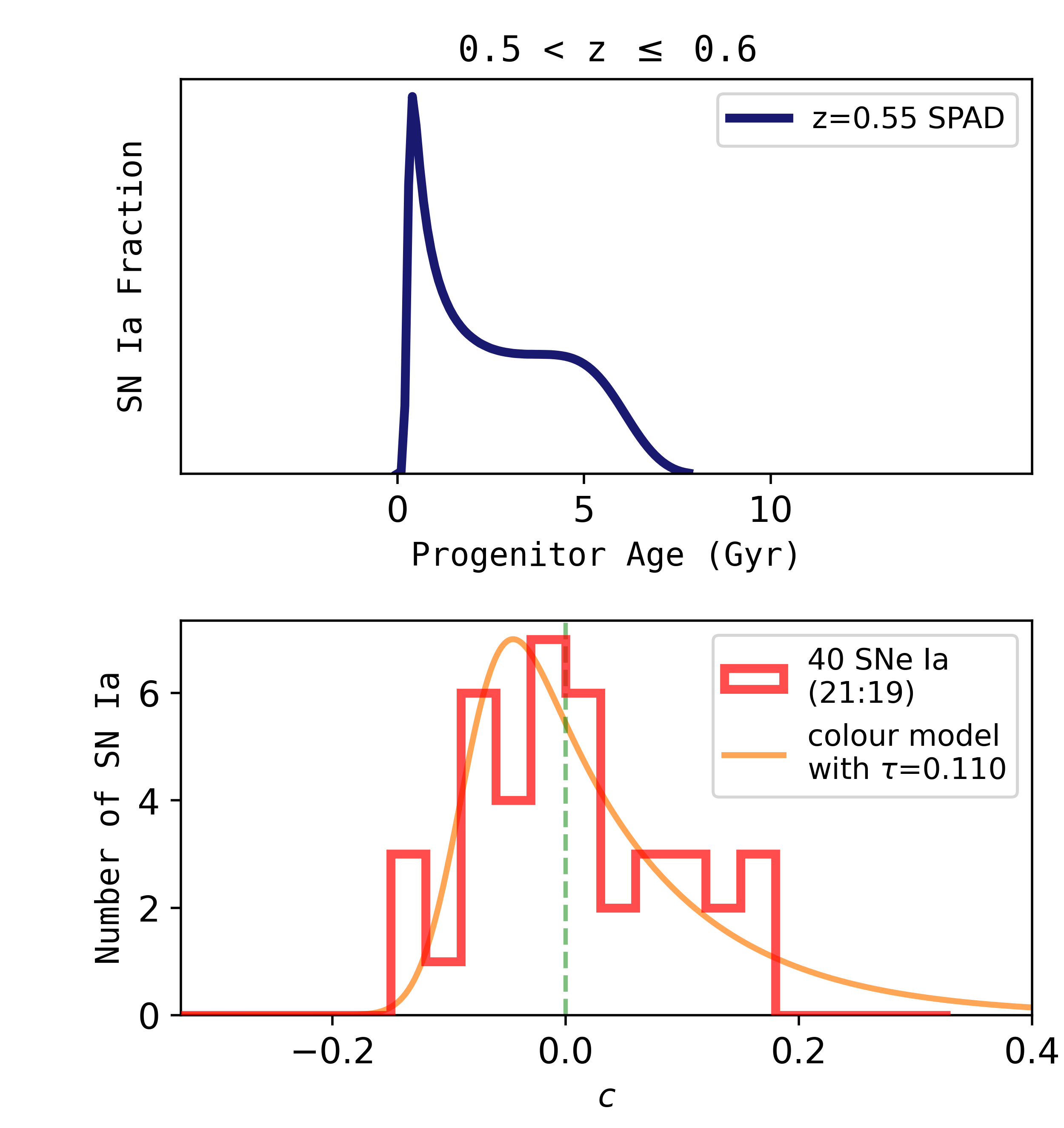}
  	\includegraphics[width=0.3\textwidth]{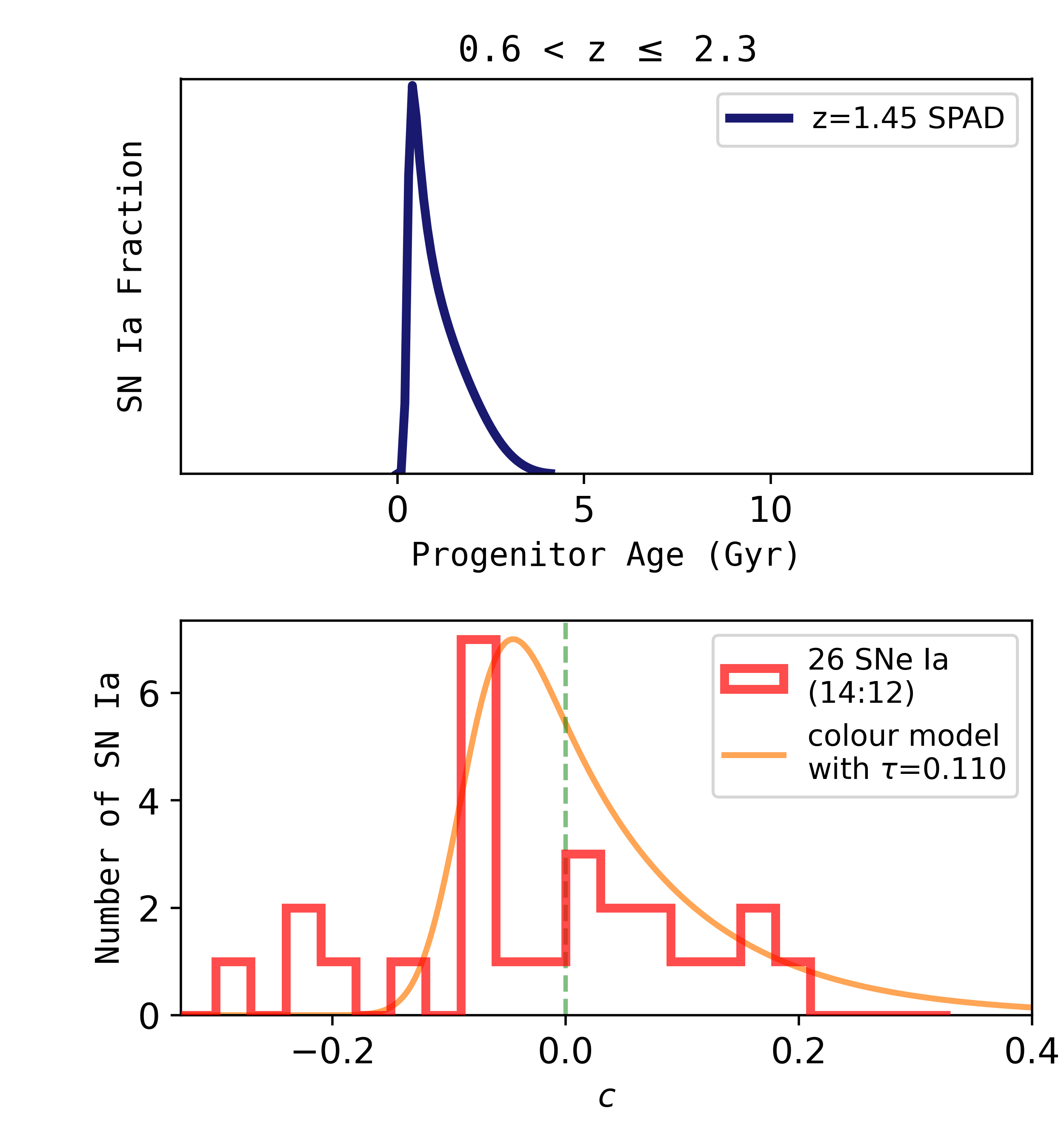}
  \caption{Same as Fig.~\ref{fig:x1_spad}, but for SALT2.4 $c$. 
  		The empirical colour model (Eq.~\ref{eq:colour_model}) is overplotted on the $c$ distribution with an orange line.
		The green dashed vertical line indicates the split point at $c = 0.0$, where the sample is divided into blue and red SNe Ia.
		The number of blue and red SNe Ia are presented in the legend for each panel.
		}
  \label{fig:c_spad}
\end{figure*}

\begin{table}
\centering
\caption{Fractions of young and old progenitor stars and high- and low-$x_1$ SNe Ia (with a total number of SNe Ia) in different redshift bins. 
Note that SPAD for the $0.6 < z \le 2.3$ bin is derived based on $z=1.45$, the middle value of that redshift bin.} 
\label{tab:age_x1_fraction}
\begin{tabular}{l c c c c}
\hline\hline\\[-0.8em]
				& SPAD		&&  \multicolumn{2}{c}{Observed $x_1$}  \\[0.15em]  \cline{2-2} \cline{4-5}
Redshift			& Young:Old 	&& High-$x_1$:Low-$x_1$ 	&($N_{\text{SNIa}}$)  \\[0.15em] 
				& [\%]		&& [\%]									 \\[0.15em]
\hline\\[-1.1em]
$0.0 < z \le 0.1$ 	& 50:50 			&& 69:31				& (1040)		\\[0.30em]
$0.1 < z \le 0.2$ 	& 56:44 			&& 80:20				& (200)			 \\[0.30em]
$0.2 < z \le 0.3$ 	& 61:39 			&& 89:11				& (88)			 \\[0.30em]
$0.3 < z \le 0.4$ 	& 67:33 			&& 79:21				& (33)			 \\[0.30em]
$0.4 < z \le 0.5$ 	& 74:26 			&& 83:17				& (29)			 \\[0.30em]
$0.5 < z \le 0.6$ 	& 80:20 			&& 98:2				& (40)			 \\[0.30em]
$0.6 < z \le 2.3$ 	& 100:0 			&& 85:15				& (26)		 	\\[0.30em]
\hline
\end{tabular}
\end{table}

\section{Results}
\label{sec:results}

In this section, we explore which SN Ia observables, $x_1$ and $c$, best indicate the age of the progenitor star qualitatively and quantitatively.
First, we perform a qualitative comparison between SPAD and the observed $x_1$ and $c$ parameters.
Then, we make a simple quantitative comparison between the fraction of young and old populations in SPAD with the high- and low-$x_1$ fractions in the observed $x_1$ distributions.
Lastly, we investigate a quantitative correlation between the determined local age around the SN Ia explosion site, as a proxy for the progenitor age, and the SN Ia observables.

\subsection{SPAD versus stretch and colour}
\label{sec:spad}

Figure~\ref{fig:x1_spad} shows distributions of observed $x_{1}$ in different redshift bins, together with SPAD we constructed with the middle value of each redshift bin. 
The empirical $x_1$($z$) model of \citetalias{Nicolas2021} (Eq.~\ref{eq:n21_x1}) is also plotted with the middle value of each redshift bin, which represents the $x_1$ distribution at a given redshift.
In the $x_1$ distribution, we split the sample into high- and low-$x_1$ SNe Ia at $x_1 = -0.75$, a high-purity threshold motivated by a two-component Gaussian-mixture analysis of the Pantheon sample (see Sec.~\ref{subsec:stretch_z}), and provide the number of each in the figure.

SPAD and the distributions of $x_{1}$ and \citetalias{Nicolas2021} $x_1$($z$) model share a common shape: a peak in the young and high-$x_{1}$ region with a tail extending to the old and low-$x_{1}$ region with a bump.
As redshift increases, the bump and the mean age of the old population become younger and eventually blend with the young population.
Likewise, on average, the number and the fraction of low-$x_{1}$ SNe Ia decrease with increasing redshift.

To make a simple quantitative comparison, we compared the integrated fraction of young and old populations in SPAD with the high- and low-$x_1$ fractions in the observed $x_1$ distributions (Tab.\ref{tab:age_x1_fraction}).
We split SPAD into young and old progenitor stars at 4.1 Gyr, where their fractions are equal (50:50) at $0.0<z\le0.1$.
For the $x_1$ distribution, as described above, we use $x_1 = -0.75$ to divide the sample into high- and low-$x_1$ groups.
The fractions of young progenitors and high-$x_1$ SNe Ia generally increase as redshift increases, while those of old progenitors and low-$x_1$ SNe Ia decrease.
Notably, the observed fraction of the low-$x_1$ SNe Ia is smaller than the old progenitor star fraction.
This apparent discrepancy is expected because the old progenitor stars (in locally old or passive environments) produce both low- and high-$x_1$ SNe Ia (see the left panel of Fig.~\ref{fig:localage}).

We also compare a distribution of observed $c$ with SPAD in Fig.~\ref{fig:c_spad}.
In the figure, we overplotted the empirical colour distribution model (Eq.~\ref{eq:colour_model}) with $\tau=0.155$ for the $z\le0.1$ bin and $\tau=0.110$ for $z>0.1$ bins.
In the $c$ distribution, we split the sample into blue and red SNe Ia at $c = 0.0$ and provide the number of each in the figure.

Unlike SPAD, the observed $c$ distribution and the empirical colour model show a peak with a tail toward redder SNe Ia, but without a distinct bump.
The fraction of red SNe Ia appears to remain above a certain level ($\sim$30\%) regardless of redshift, whereas the fraction of old SN Ia progenitor stars in SPAD decreases with redshift.


\subsection{Local age versus stretch and colour}
\label{sec:local_age}

\begin{figure*}
	\centering	
  	\includegraphics[width=0.7\textwidth]{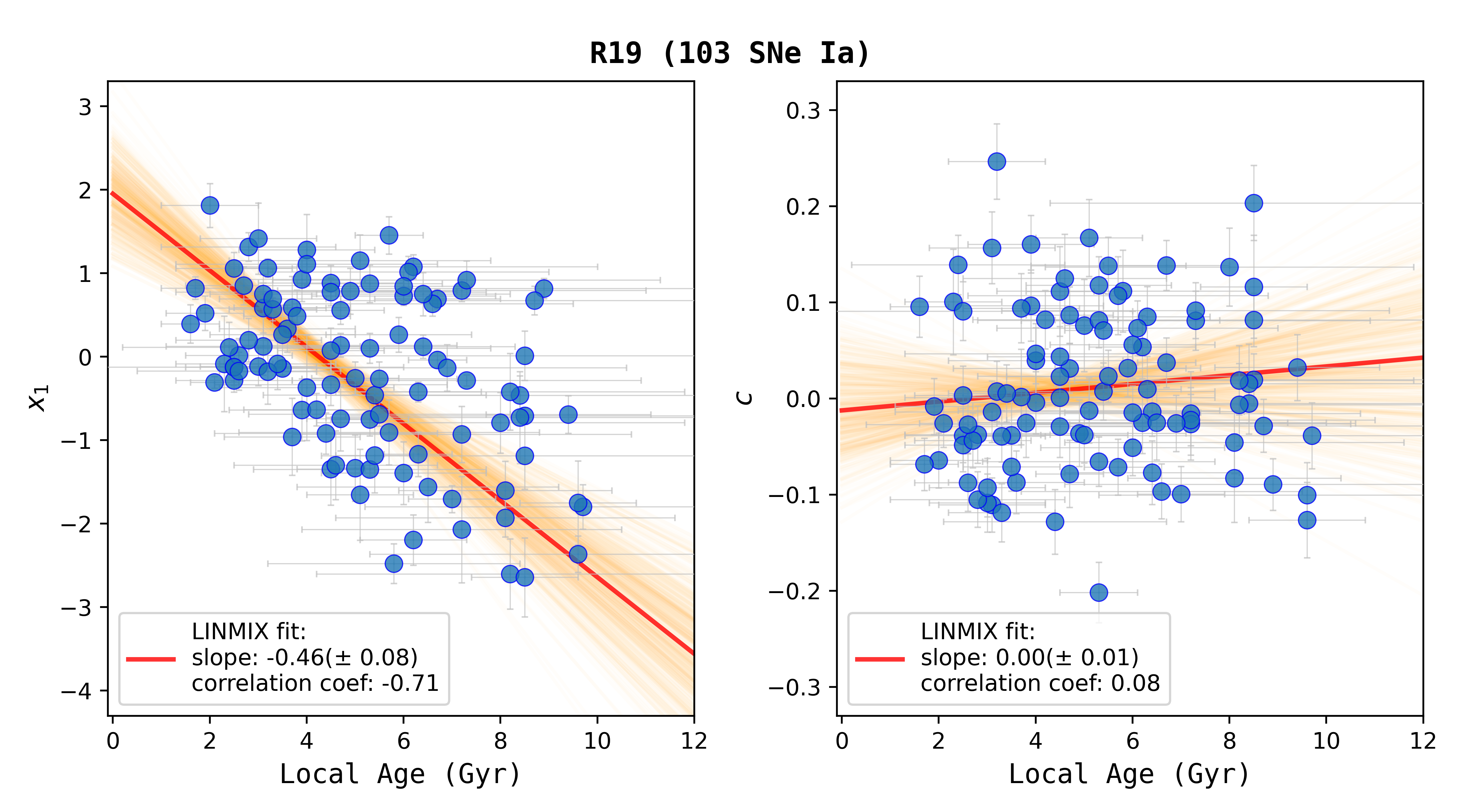}
  \caption{Local age versus SALT2 $x_{1}$ (left panel) and $c$ (right) taken from \citetalias{Rose2019}.
  		A red solid line shows the average of 10,000 linear regression results (light red lines) from the LINMIX package, which returns a slope of $-0.46 \pm 0.08$ between local age and $x_1$, and of $0.00 \pm 0.01$ between local age and $c$.
		Linear correlation coefficients estimated by the LINMIX are also provided in each panel.
		}
  \label{fig:localage}
\end{figure*}

\begin{figure*}
	\centering	
  	\includegraphics[width=0.7\textwidth]{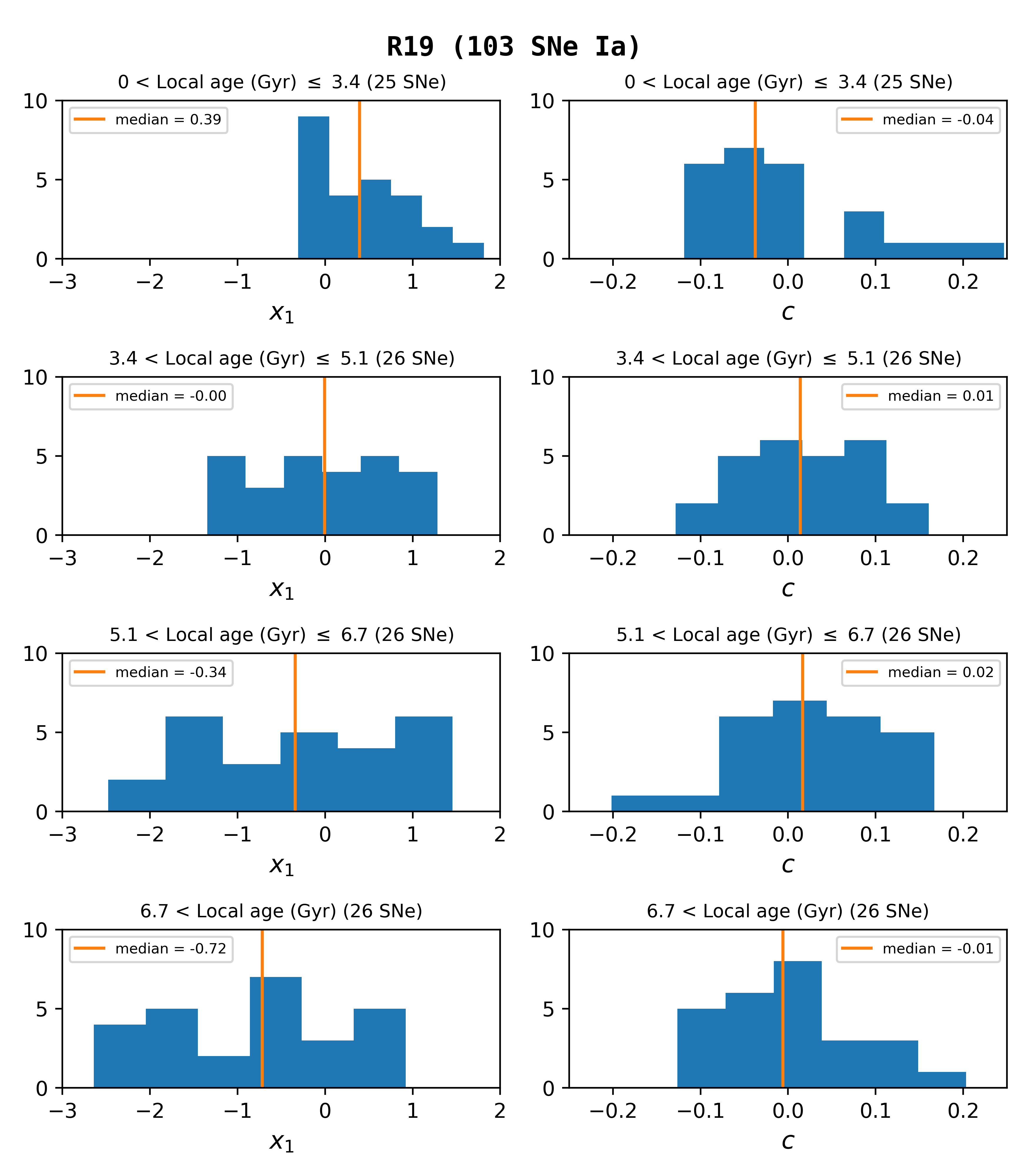}
  \caption{Distributions of SALT2 $x_{1}$ (left panel) and $c$ (right) split by the local age that has the same sample size.
		Orange vertical lines represent median values for each distribution.
		}
  \label{fig:localage_split}
\end{figure*}

\begin{figure}
	\centering	
  	\includegraphics[width=0.45\textwidth]{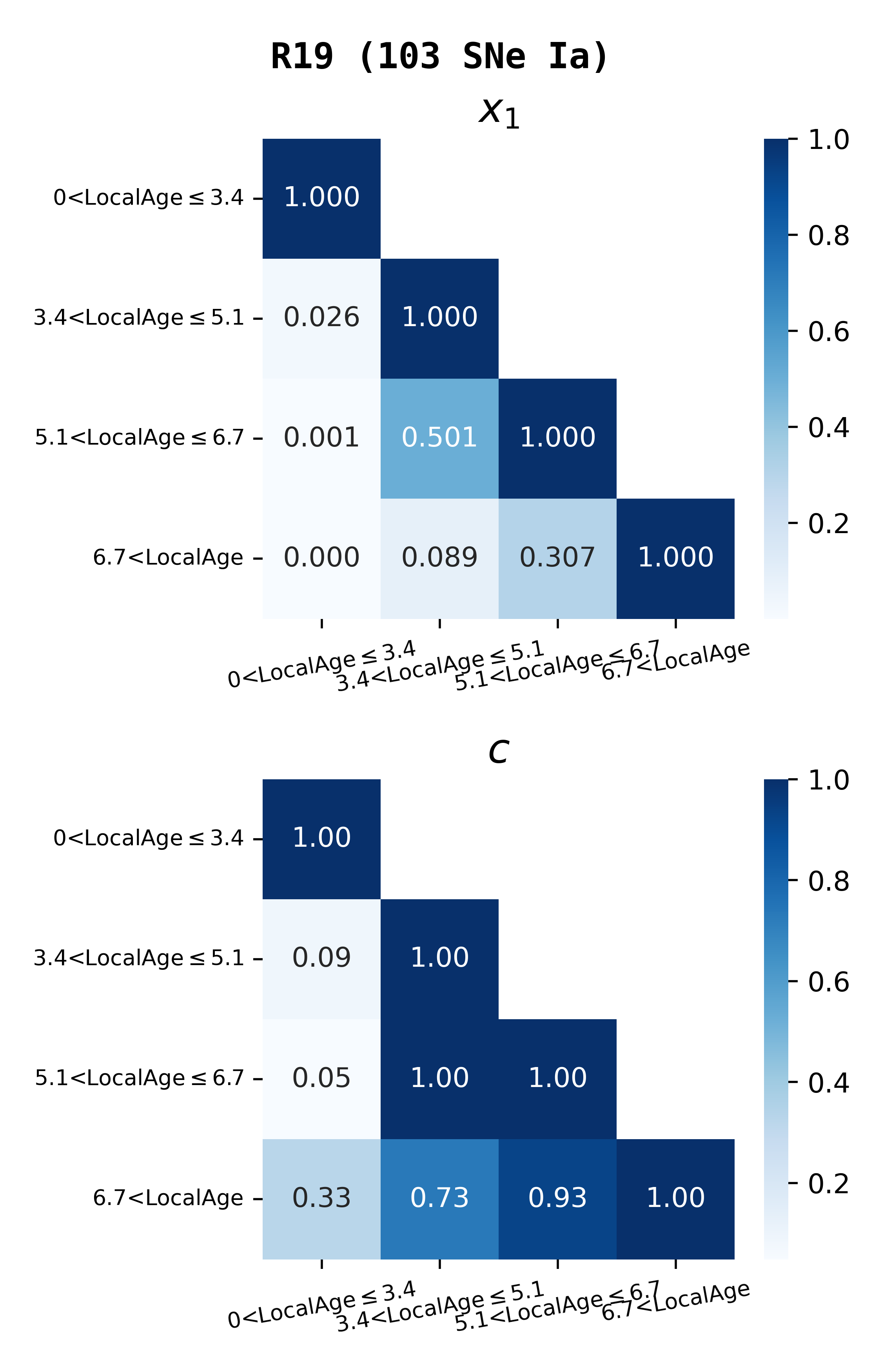}
  \caption{$p$-values of the two-sample Kolmogorov-Smirnov test on the $x_1$ (top panel) and $c$ (bottom) distributions in different local age bins of Fig.~\ref{fig:localage_split}.
  		Note that the real value of $0.000$ in the $x_1$ plot is $5.41\times10^{-6}$.
		}
  \label{fig:localage_pvalue}
\end{figure}

There is no direct way to investigate a one-to-one correlation between SPAD and stretch and colour, because no progenitor star age has been determined thus far.
However, there is an indirect approach that uses local environment information around the SN Ia explosion site, e.g., local age, assuming this age traces the progenitor star age.
Previous studies used environmental proxies for the local age, such as the local star formation rate \citep{Rigault2013, Rigault2020} or local colour \citep{Roman2018, Kelsey2021}.
They showed that SNe Ia in locally star-forming or blue environments preferentially have high-stretch values, while the colour is independent of local environments.
Since the star formation rate, colour, age, and metallicity are broadly correlated, one can interpret these results in the context of age.
Unlike previous studies, in this work, we utilize the explicitly determined local age around the SN Ia explosion site adopted from \citet[hereafter \citetalias{Rose2019}]{Rose2019}.

\citetalias{Rose2019} provide the local age for 103 SNe Ia along with SALT2 parameters.
Briefly, they selected SNe Ia from SDSS-II SN survey that passed cosmological cuts as described and released by \citet{Campbell2013} and did local $ugriz$ photometry within a radius of 1.5 kpc of the SNe Ia location.
Then, they used a Bayesian method to determine local environment properties including mass--weighted age by matching observed spectral energy distributions to a synthesized stellar population model of Flexible Stellar Population Synthesis code \citep{Conroy2009, Conroy2010}.
We refer the reader to \citetalias{Rose2019} for more detailed description of the estimation of the local environment properties.

With the \citetalias{Rose2019} sample, we plot in Fig.~\ref{fig:localage} correlations of local age with SALT2 $x_1$ and $c$.
We overplotted 10,000 linear regressions on the \citetalias{Rose2019} sample using the LINMIX package in \texttt{Python}\footnote{\href{https://github.com/jmeyers314/linmix/}{https://github.com/jmeyers314/linmix/}}, which employs an MCMC posterior sampling and a hierarchical Bayesian approach, considering errors in both variables \citep{Kelly2007}.
The LINMIX returns a slope and a linear correlation coefficient.
Between local age and $x_1$, the slope is $-0.46 \pm 0.08$ and the linear correlation coefficient is $-0.71$, indicating they are well correlated.
However, between local age and $c$, the slope is $0.00 \pm 0.01$ and the linear correlation coefficient is $0.08$, which shows no correlation.
Note that we also see some high-$x_1$ SNe Ia in locally old environments, as observed in \citet{Lee2022} using the global age of host galaxies.
This shows that SNe Ia in locally old environments contain a mixture of $x_1$ values, and therefore caution is needed when using the $x_1$--local age correlation to infer the (mean) $x_1$ at a given age, and vice versa.

Instead, to examine how the distributions of $x_1$ and $c$ change with the local age, Fig.~\ref{fig:localage_split} presents their distributions split by the local age. 
The sample size for each local age bin is 26, but the first bin has 25 SNe Ia.
The $x_1$ distribution systematically evolves with the local age, such that the youngest environments have only high-$x_1$ ($> -0.75$) SNe Ia and the number of low-$x_1$ SNe Ia increases with the local age.
Therefore, the median value of the $x_1$ distribution varies from $+0.39$ to $-0.72$.
In contrast, the $c$ distribution and its median value do not vary with the local age.

Additionally, we performed the two-sample Kolmogorov-Smirnov (KS) test on the $x_1$ and $c$ distributions in different local age bins of Fig.~\ref{fig:localage_split}.
Fig.~\ref{fig:localage_pvalue} shows the $p$-values for each combination.
The $x_1$ distribution at the youngest local age bin differs significantly from those in the other local age bins, with $p$-values $\le0.026$.
Notably, the $p$-value between the $x_1$ distributions at the youngest and the oldest local age bins is a very low value of $5.41\times10^{-6}$.
However, $c$ distributions show no significant difference between the local age bins as suggested by $p$-values ($\geq 0.05$).

\section{Empirical mapping from SPAD to the $x_1$ distribution}
\label{sec:forward_model}

\begin{figure*}
  	\includegraphics[width=0.3\textwidth]{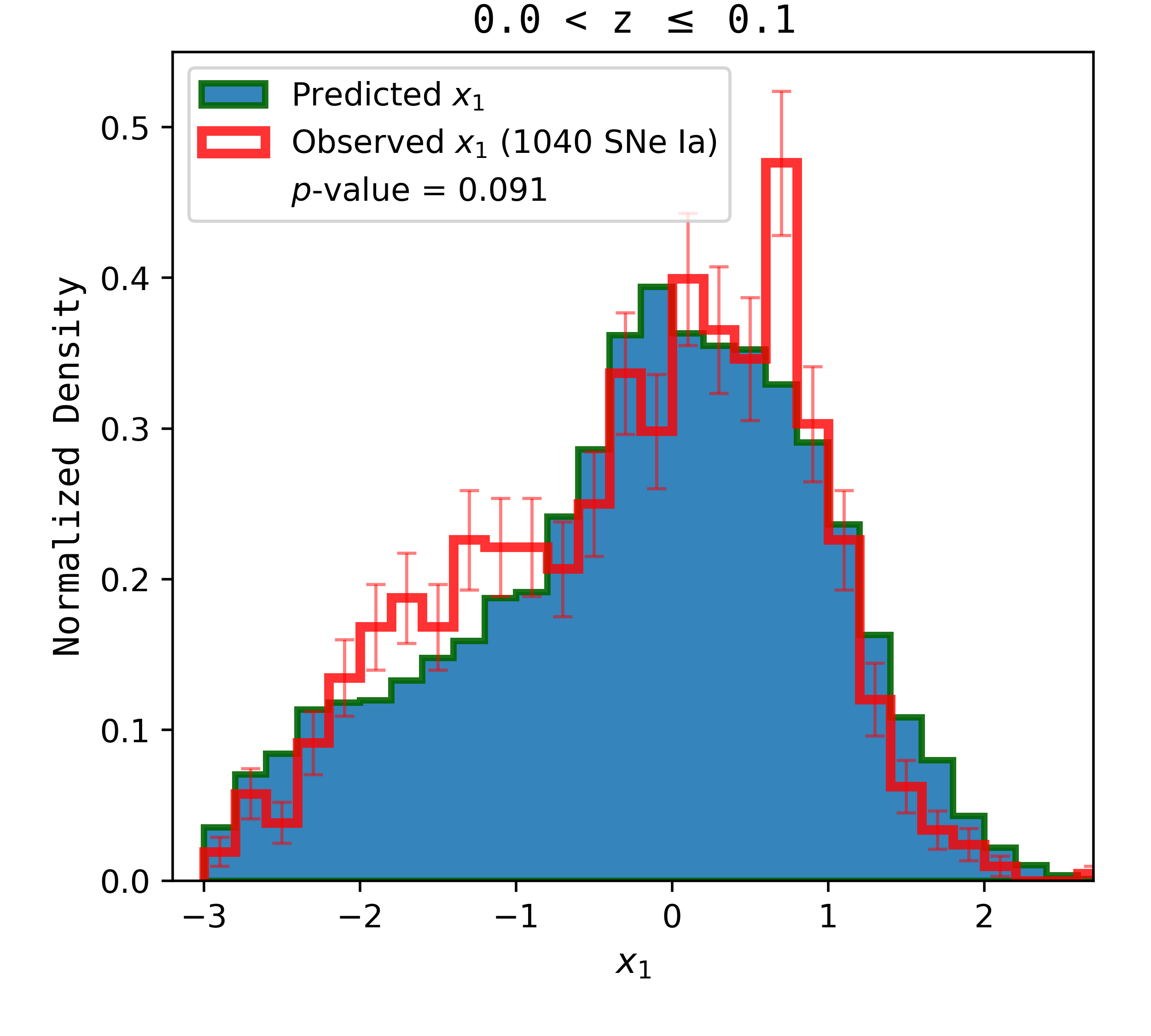}
  	\includegraphics[width=0.3\textwidth]{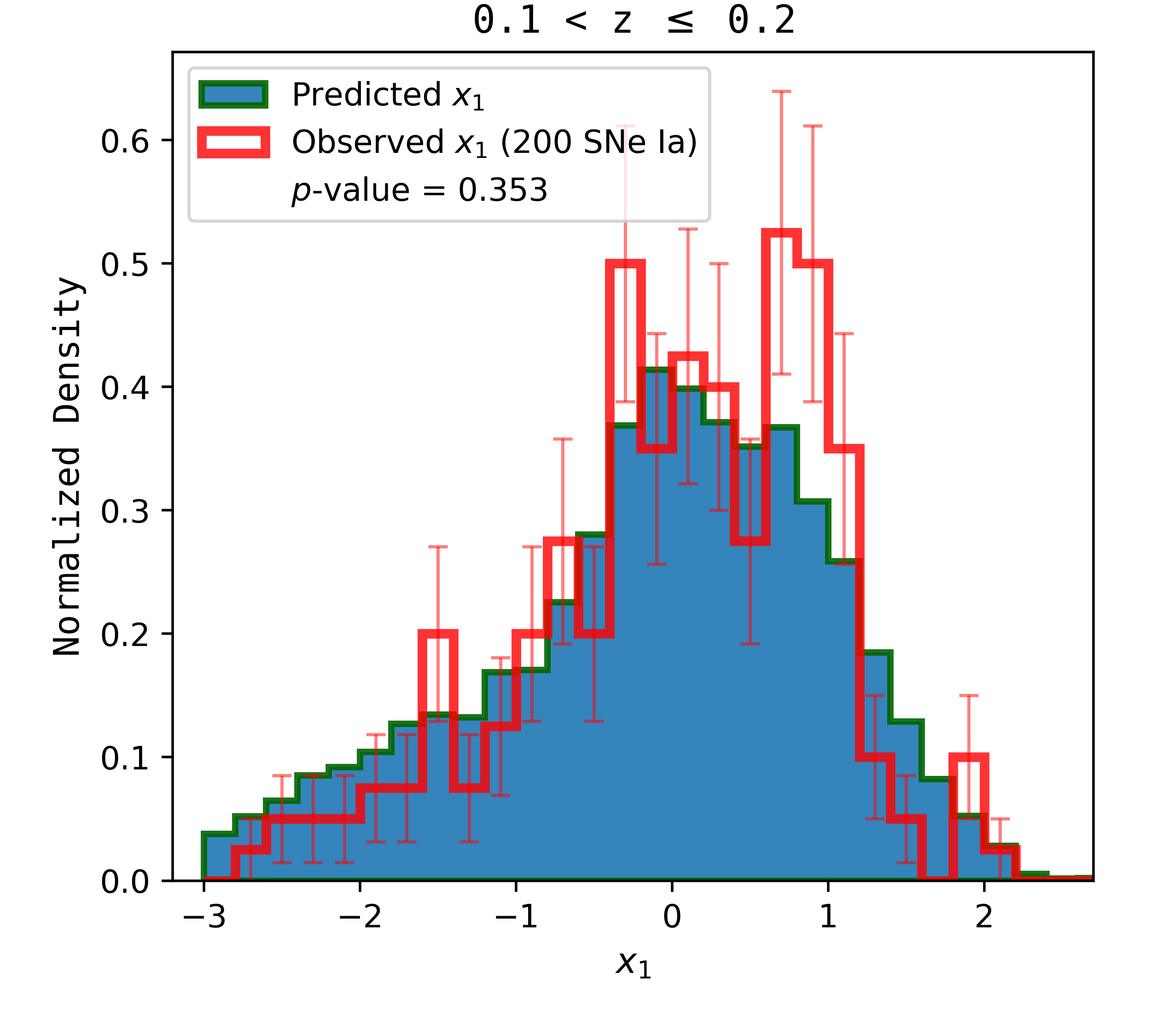}
  	\includegraphics[width=0.3\textwidth]{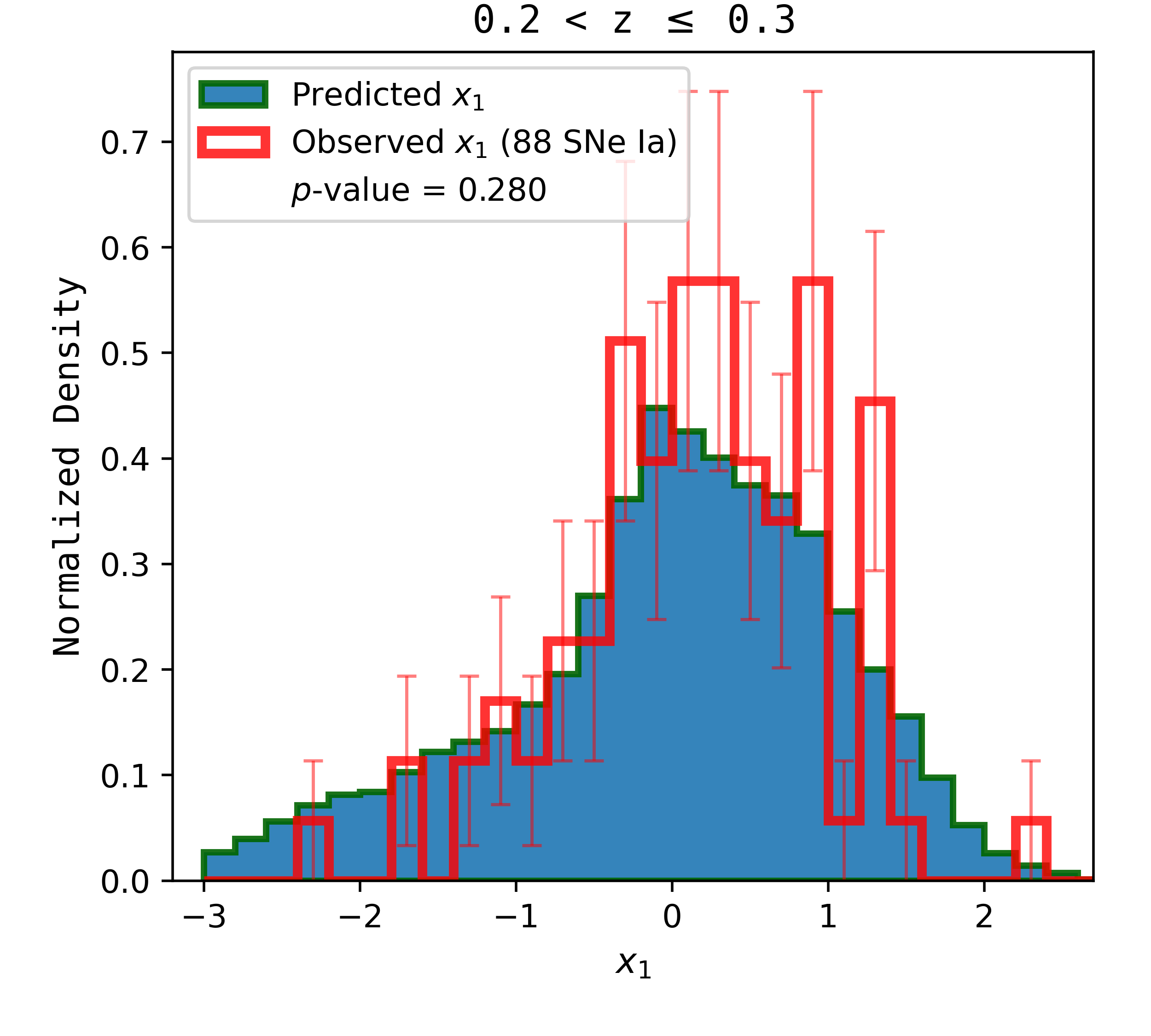}
  	\includegraphics[width=0.3\textwidth]{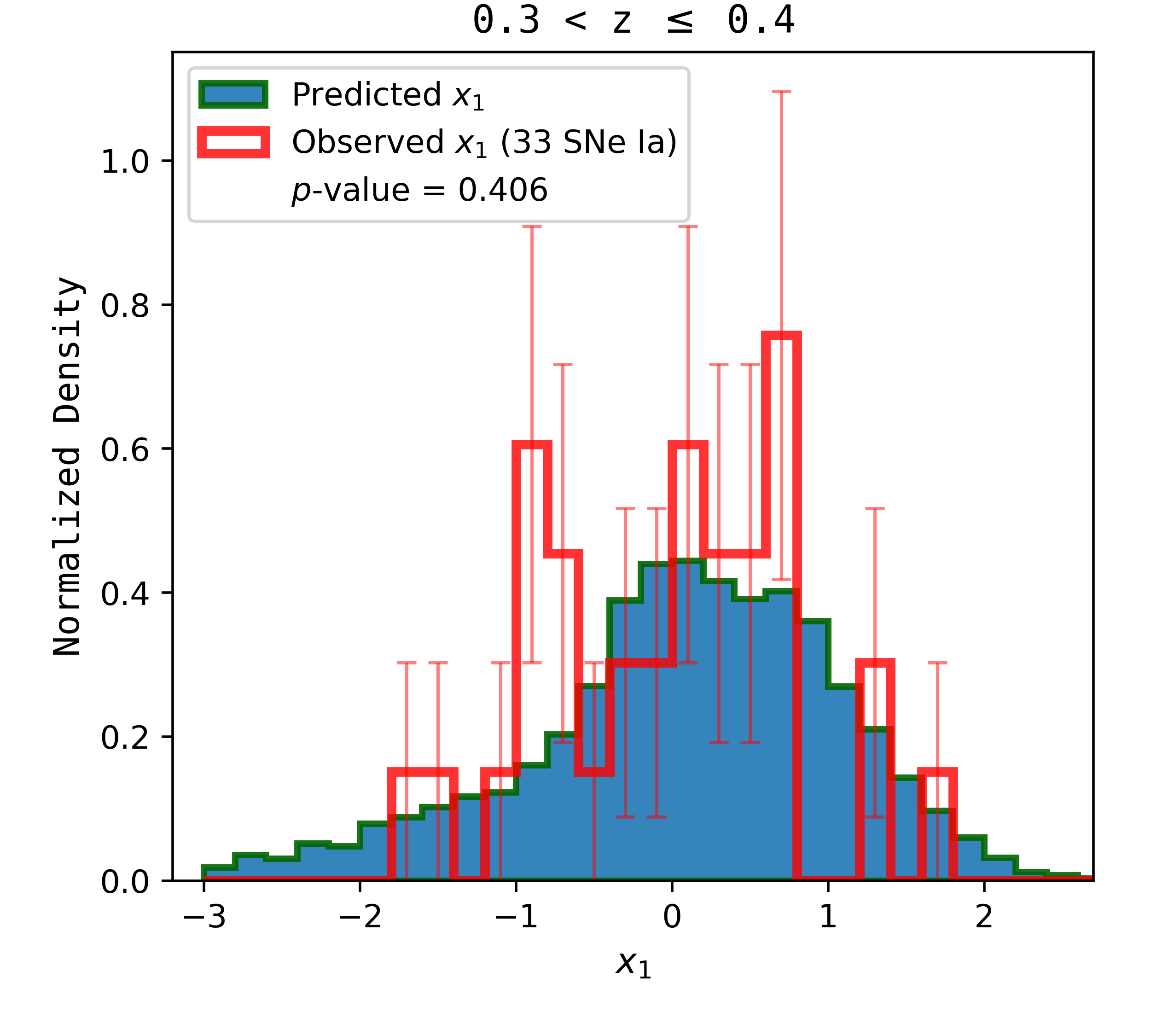}
  	\includegraphics[width=0.3\textwidth]{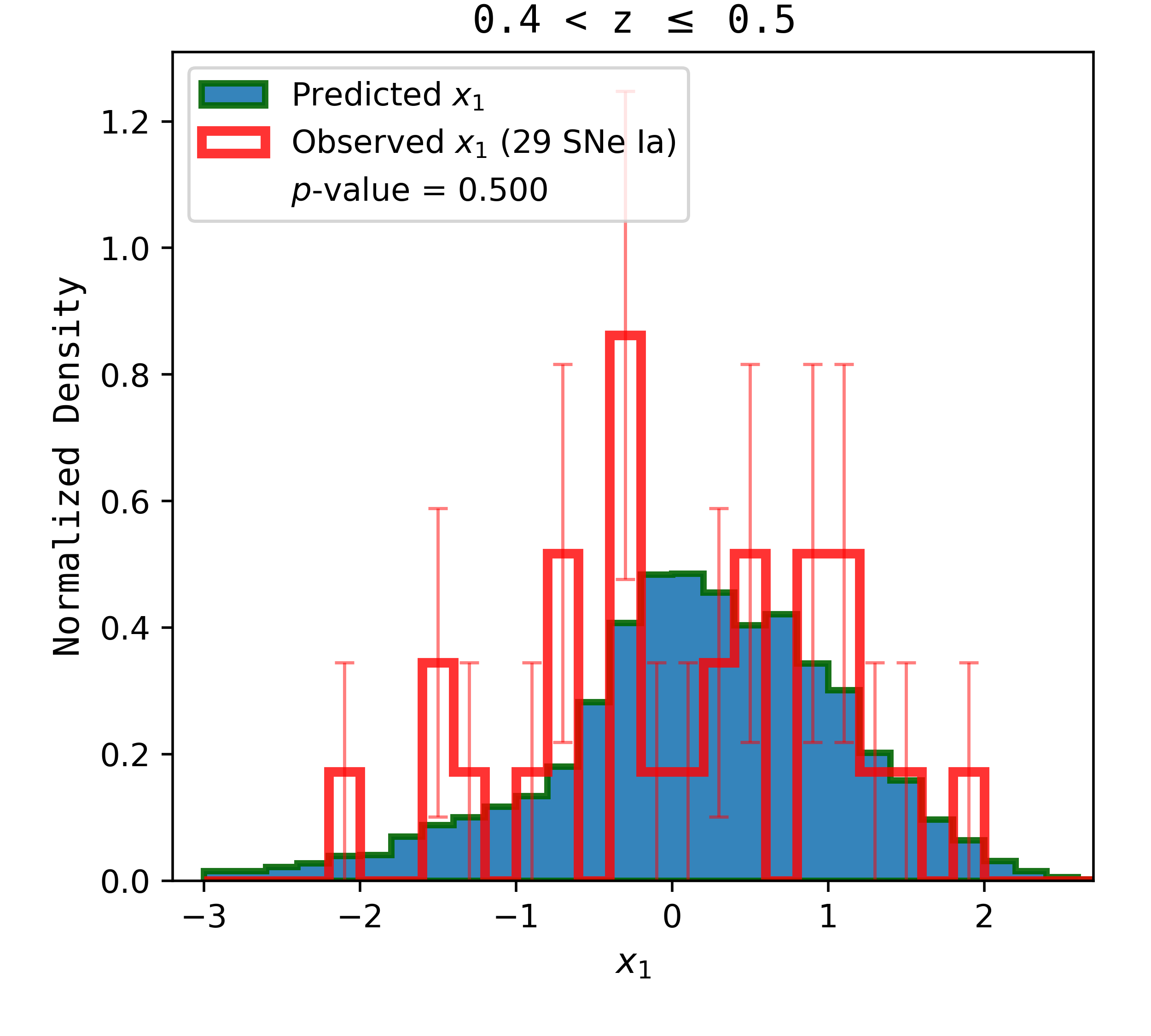}
  	\includegraphics[width=0.3\textwidth]{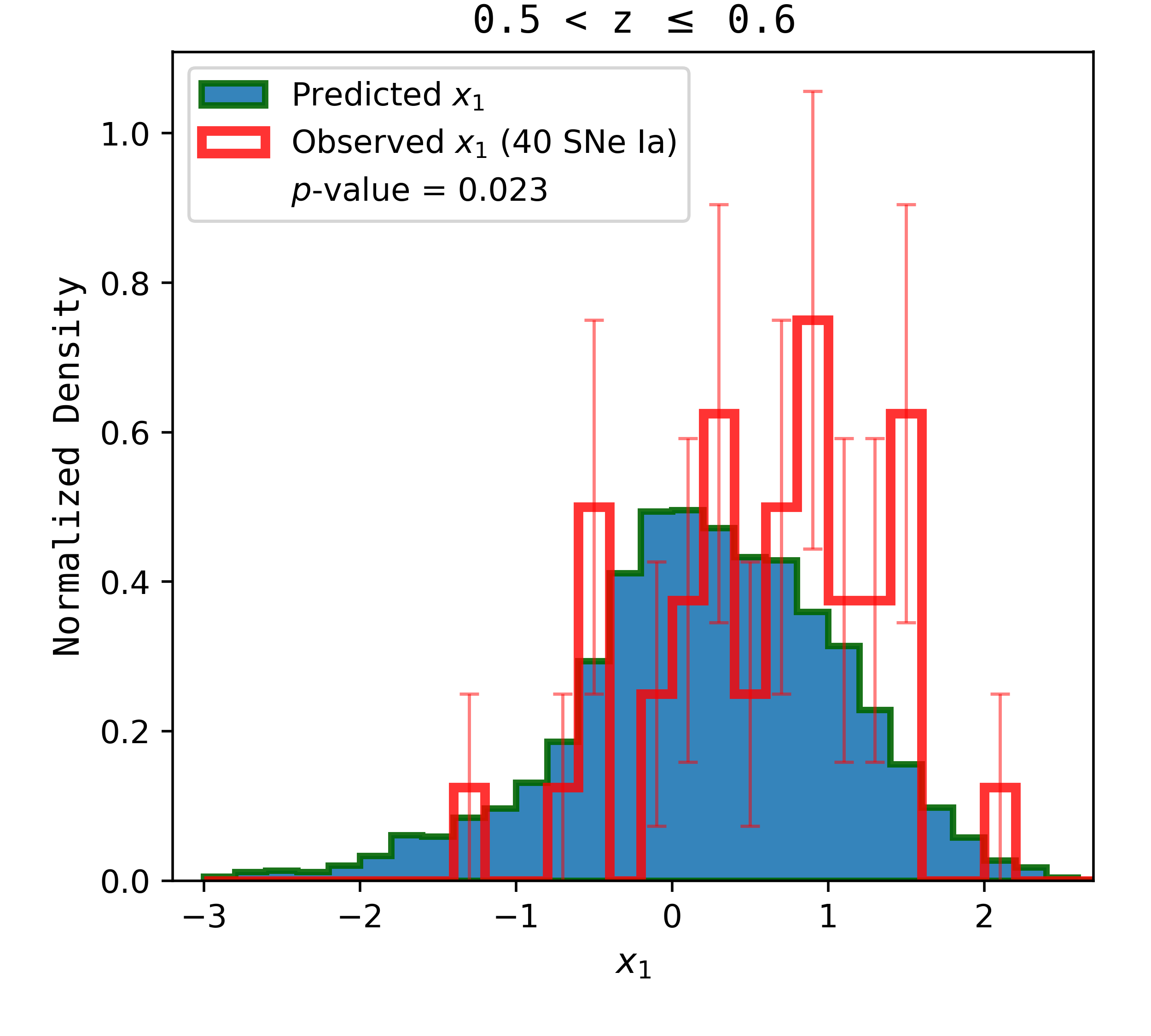}
  	\includegraphics[width=0.3\textwidth]{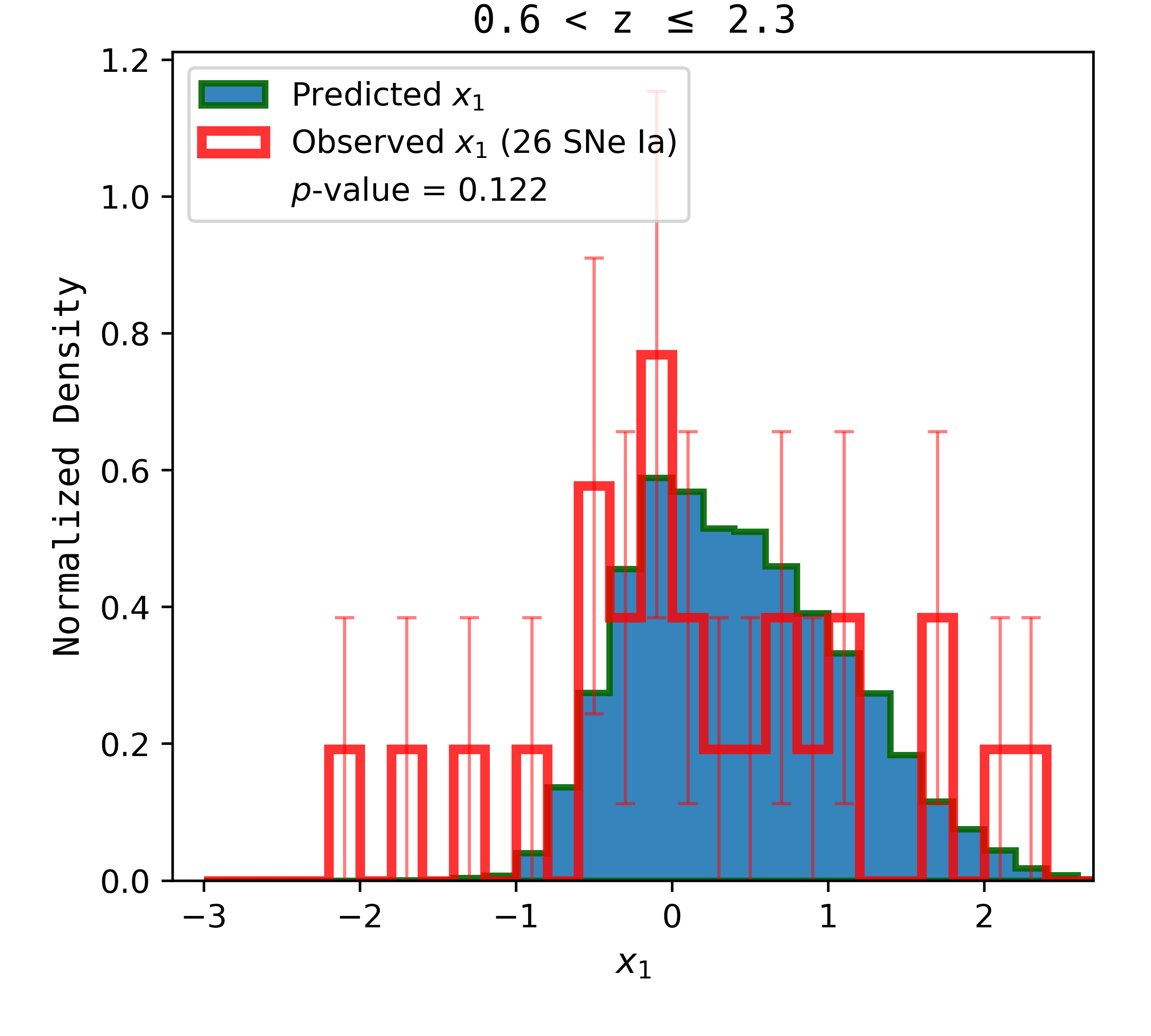}
  \caption{Comparison of predicted and observed $x_1$ distributions in different redshift bins.
  		Blue filled histograms show the predicted $x_1$ distribution from our empirical mapping approach based on SPAD, while red step histograms show observed $x_1$ distribution for our sample.
		For the observed $x_1$, Poisson statistical errors are overlaid on each bin.
		$p$-values of the two-sample KS test are presented in the label.
		}
  \label{fig:forward_model}
\end{figure*}

Based on the above qualitative and quantitative comparisons, we argue that the light-curve shape distribution indicates the age of the SN Ia progenitor star.
To validate this argument, we employ a simple empirical mapping approach to predict the observed $x_1$ distribution from SPAD.
Different from \citet{Wiseman2022}, who employed a more complicated forward modelling approach to simulate the $x_1$ distribution by combining a galaxy-driven model with the empirical $x_1$ prescription of \citetalias{Nicolas2021}, our approach maps SPAD directly to $x_1$ using the empirically determined local age--$x_1$ relationship we analyzed (Sec.~\ref{sec:local_age}), without assuming a parametric functional form for the $x_1$ distribution.
This provides a more physically direct link between SPAD and the observed light-curve shape.

For a given redshift, we first drew 100,000 progenitor ages from SPAD to generate a mock SN Ia sample.
We then mapped each progenitor age to an $x_1$ value using the empirical local age--$x_1$ relation determined in Sec.~\ref{sec:local_age}.
In this process, we divided the empirical local age--$x_1$ data into four distinct age bins, following the same four age bins defined in Sec.~\ref{sec:local_age}, to capture the physical scatter, such as the presence of high-$x_1$ SNe in old environments (see Fig.~\ref{fig:localage_split}).
For each age bin, we constructed a non-parametric Gaussian Kernel Density Estimation (KDE) to capture the empirical probability density of the $x_1$ distribution.
The KDE bandwidth was determined using Scott's rule, which scales with the standard deviation and sample size of the input data, and thus scatter in $x_1$ at a given progenitor age is naturally incorporated without additional assumptions.
Each mock SN was then assigned an $x_1$ value sampled from the KDE of its corresponding age bin.
Finally, to ensure a fair comparison with the observed data, we applied the same selection cut ($-3 < x_1 < 3$) as in the observed data to our mock $x_1$ distribution.
Fig.~\ref{fig:forward_model} shows the resulting predicted $x_1$ distributions alongside the observed ones.

Our empirical mapping approach appears to successfully reproduce the observed $x_1$ distribution in different redshift bins, given that the apparent differences between the predicted and observed distributions are well within the Poisson statistical errors.
To quantify this, we performed the two-sample KS test.
The KS test results at $z \le 0.3$, where the sample size is sufficiently large, show no statistically significant difference between our predicted and observed $x_1$ distributions with $p$-values of 0.09, 0.35, and 0.28 for $0.0<z\le0.1$, $0.1<z\le0.2$, and $0.2<z\le0.3$, respectively.
This lack of significant deviation was consistently observed in higher redshift ranges, despite the smaller sample sizes.

\section{Discussion}
\label{sec:discussion}

The main advance of this work is not simply to re-examine the observed SN Ia light-curve shape and colour distributions, but to test them against an explicitly predicted progenitor star age structure and to validate that interpretation with local age measurements and the empirical mapping approach.
We construct SPAD through multiplying the SN Ia DTD model by the collected cosmic SFHs following the \citetalias{Childress2014} framework.
From the qualitative comparison of SPAD with the observed distributions of SALT2.4 $x_1$ (the light-curve shape) and $c$ (the colour), we find that SPAD and the $x_1$ distribution share a common shape: a young/high-$x_1$ peak and an old/low-$x_1$ bump in the tail, and this shape varies systematically with redshift. 
In contrast, this trend is not evident in the $c$ distribution.
Then, the quantitative comparison of the local age at the SN Ia explosion site, used as a proxy for the age of the SN Ia progenitor star, with the observed distribution of $x_1$ and $c$ shows that $x_1$ is well correlated with the local age (the linear correlation coefficient $\simeq -0.71$), while there is no correlation with $c$ (the coefficient $\simeq 0.08$).
Furthermore, we find that the $x_1$ distribution systematically evolves with the local age.
Notably, the $x_1$ distribution at the youngest local age bin differs significantly from those in the other local age bins, with $p$-values $\le0.026$.
In contrast, regarding the $c$ distribution, we find no evolution and no significant difference between the local age bins ($p$-values $\geq 0.05$).
Finally, our empirical mapping approach based on SPAD successfully reproduces the observed $x_1$ distribution across different redshift bins.
Taken together, our results suggest that the light-curve shape distribution indicates progenitor star age at the population level more robustly than the colour does.

In this section, we discuss an application regarding the Malmquist bias in SN surveys, a homogeneous SN Ia sample, and future works based on our finding.

\subsection{Only high-stretch and young progenitor SNe Ia in the higher redshift?: the Malmquist bias or the intrinsic SN Ia property? }
\label{subsec:stretch_z}

\begin{figure*}
  \centering
  \includegraphics[width=\columnwidth]{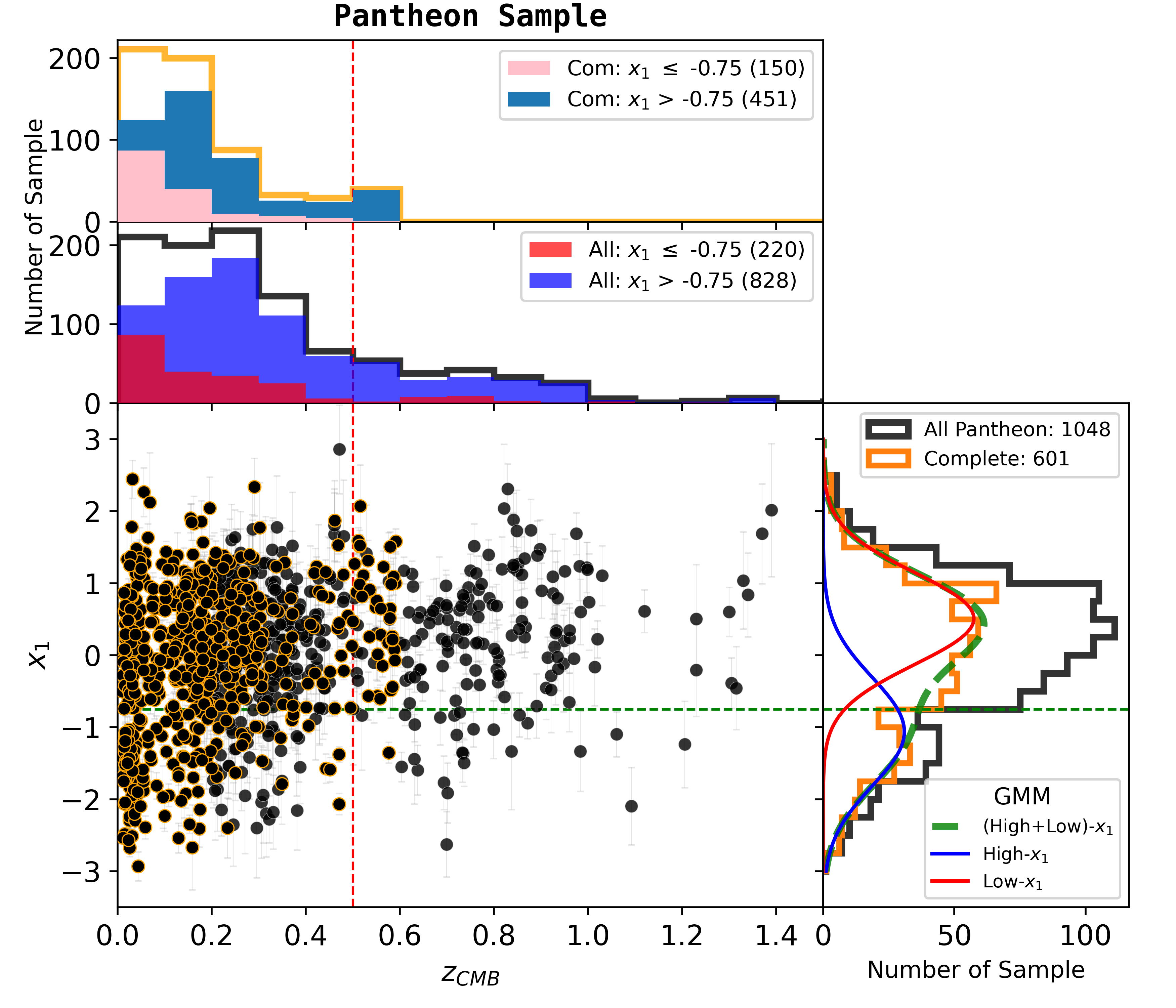}
  \includegraphics[width=\columnwidth]{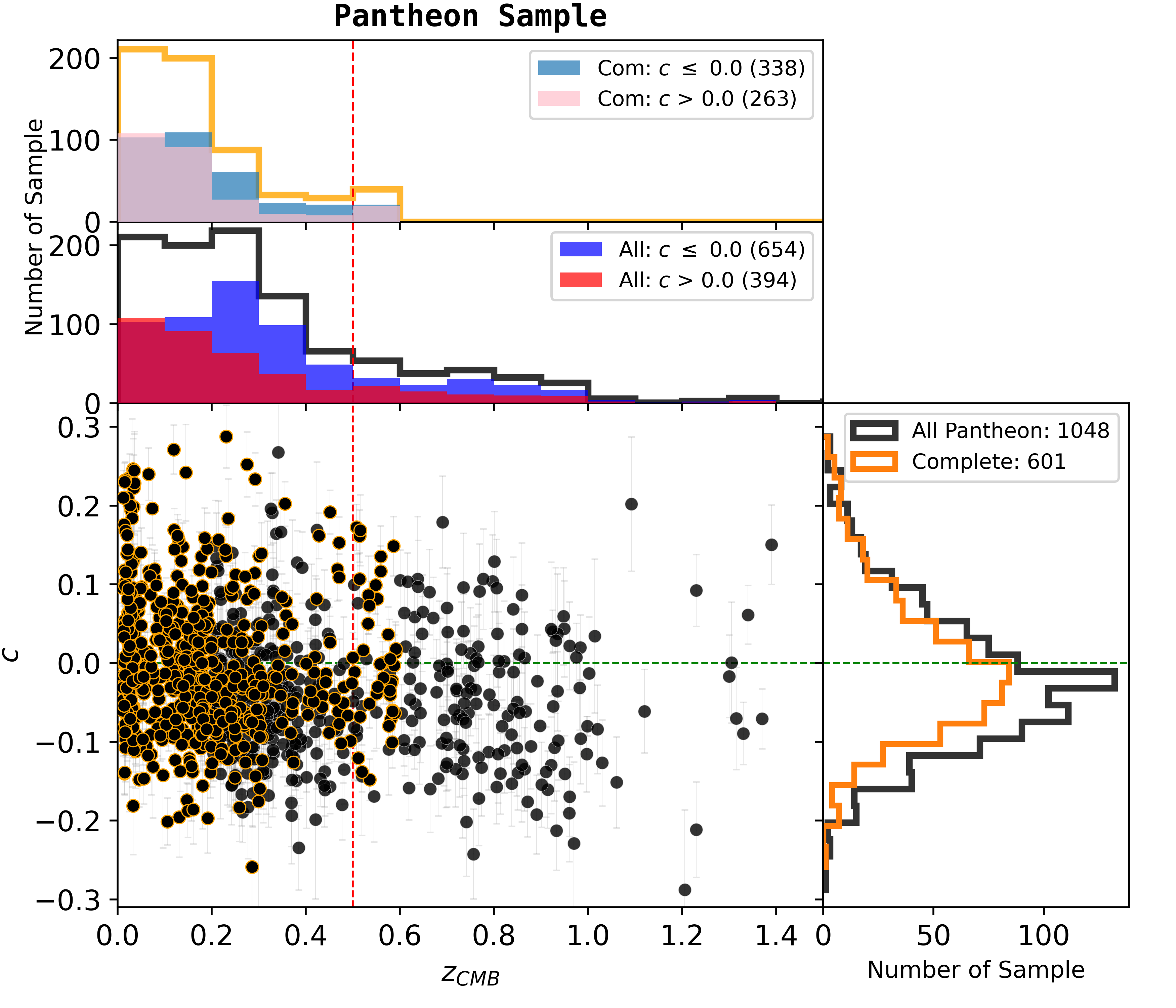}
  \caption{Distributions of SNe Ia $x_1$ (left main panel) and $c$ (right main panel) as a function of redshift for all Pantheon (black marks) and volume-limited Pantheon (orange solid line around the black marks; denoted as `Com' or `Complete' in the legends) samples.
  		In each main panel, the right histogram shows a $x_1$ or $c$ distribution, and the top histogram presents a redshift distribution of the sample.
		Histograms are separated by different colours for all Pantheon and volume-limited Pantheon samples, as indicated in the labels.
		In the left main panel, based on the $x_1$ histogram and the two-component Gaussian mixture model (GMM) (bottom right sub-panel), we divide the Pantheon sample into high- and low-$x_1$ SNe Ia at $x_1 = -0.75$ (green dashed horizontal line) to represent a high-purity ($>93\%$) low-$x_1$ sample.
		In the main right panel, we split the sample into red and blue SNe Ia with $c = 0.0$ (green dashed horizontal line).
		Red dashed vertical lines in each panel indicate $z = 0.5$, where the passive sequence of galaxies begins to appear.
		}
  \label{fig:z_vs_x1c}
\end{figure*}

\begin{table*}
\centering
\caption{Number of low-/high-$x_1$ and red/blue SNe Ia in all Pantheon and volume-limited Pantheon sample splitting at $z=0.5$.}
\label{tab:z_x1c}
\begin{tabular}{l c c |  c c |  c}
\hline\hline\\[-0.8em]
\multicolumn{6}{c}{All Pantheon Sample (1048 SNe Ia)} \\[0.15em]
\hline
		& Low-$x_1$ 		& High-$x_1$ 		& Red 		& Blue 		& \multirow{2}{*}{Total} \\[0.15em] 
		& ($x_1 \le -0.75$) 	& ($x_1 > -0.75$) 	& ($c > 0.0$) 	& ($c \le 0.0$)	& \\[0.15em] 		
\hline\\[-1.1em]
$z > 0.5$ 	& 27 (12.5\%) 		& 189 			& 77 (35.6\%) 	& 139 		& 216 \\[0.30em]
$z \le 0.5$ & 193 (23.2\%) 	& 639 			& 317 (38.1\%)	& 515 		& 832 \\[0.30em]
\hline
Total		& 220 (21.0\%) 		& 828 			& 394 (37.6\%)	& 654 		& 1048 \\
\hline
\\
\hline \hline \\[-0.8em]
\multicolumn{6}{c}{Volume-Limited Pantheon Sample  (601 SNe Ia)} \\[0.15em]
\hline\\[-1.1em]
		& Low-$x_1$		& High-$x_1$		& Red		& Blue		& Total \\[0.15em]
\hline\\[-1.1em]
$z > 0.5$ 	& 1 (2.5\%) 		& 39 				& 19 (47.5\%) 	& 21 			& 40 \\[0.30em]
$z \le 0.5$	& 149 (26.6\%) 		& 412 			& 244 (43.5\%)	& 317 		& 561 \\[0.30em]
\hline
Total		& 150 (25.0\%) 		& 451 			& 263 (43.8\%)	& 338 		& 601 \\
\hline
\end{tabular}
\end{table*}

\begin{figure}
  \centering
  \includegraphics[width=\columnwidth]{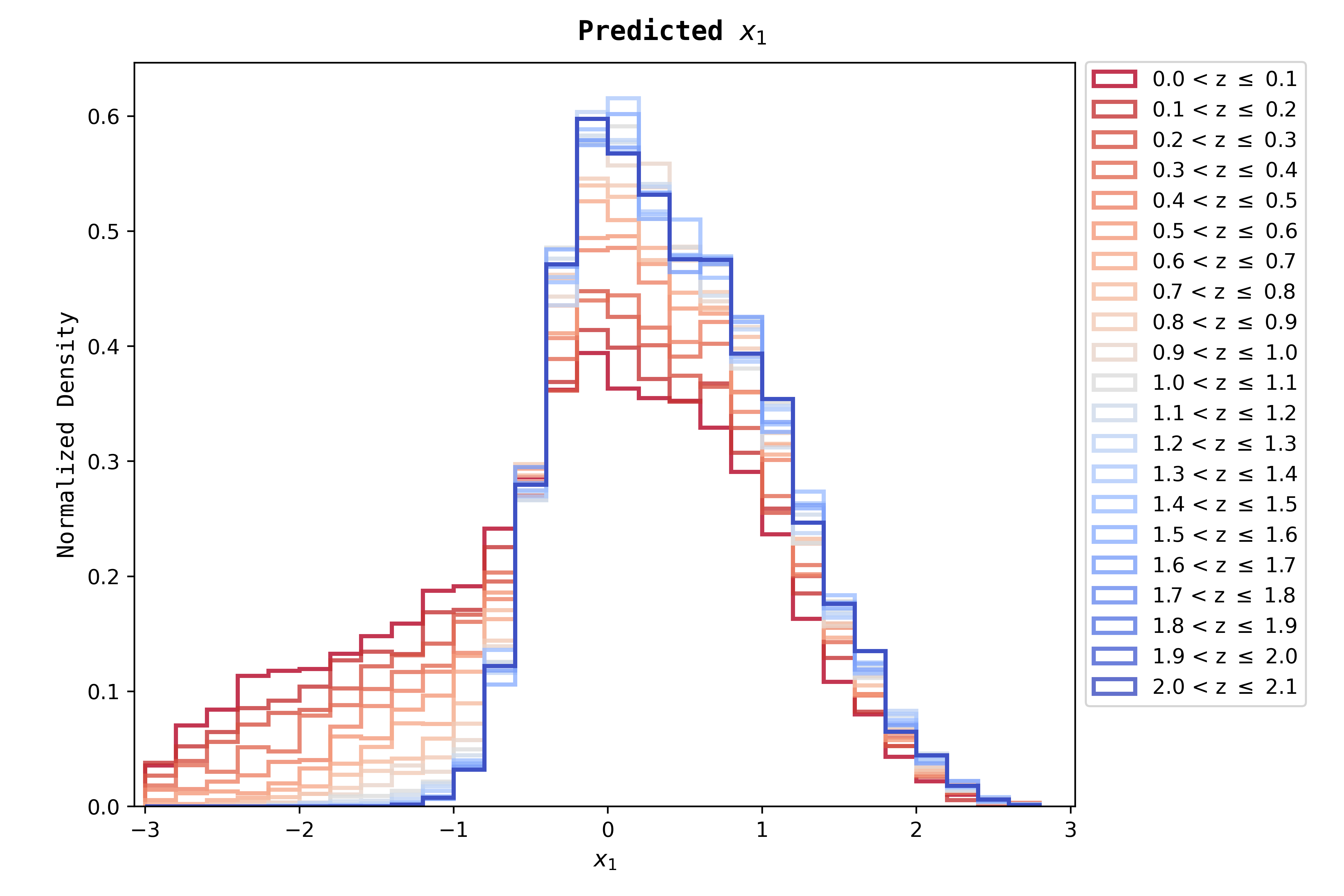}
  \caption{Distributions of SN Ia $x_1$ in different redshift bins predicted from our empirical mapping approach.
  		The colour of the histogram changes from red to blue as redshift increases.
		}
  \label{fig:pred_x1}
\end{figure}

Previous host galaxy studies established that low-$x_1$ SNe Ia prefer to explode in the passive environments, while $c$ is independent of the star formation environment \citep{Sullivan2010, Rigault2013, Pan2014, Roman2018, Kim2019, Rigault2020, Kelsey2021}.
In the galaxy evolution studies, they showed that a passive sequence of galaxies begins to appear clearly at $z\sim0.5$ \citep[see][and references therein]{Sparre2015, Rodriguez-Puebla2017, Moster2019}.
Based on these collective results, we hypothesize that SNe Ia with low-$x_1$ value, which indicates the old progenitor star, would similarly begin to be observed at $z\sim0.5$.
This hypothesis is consistent with our SPAD, where the peak of older progenitor star becomes significant at $z\le0.5$ (see Figure~\ref{fig:spad_z}).
In contrast, since the SN Ia colour is independent of the star formation environment, the change in the fraction of this environment across $z\sim0.5$ should leave the fraction of red SNe Ia unchanged, unlike that of low-$x_1$ SNe Ia\footnote{We note that this prediction is distinct from the change in the colour distribution at $z\sim0.1$. This change results from the decrease of the extrinsic dust component, whose evolution is probed by \citet{Popovic2025}, not from a change in the intrinsic SN Ia colour, and occurs well below the $z\sim0.5$ boundary considered here.}.
In order to test our predictions, we plot $x_1$ and $c$ as a function of redshift in Fig.~\ref{fig:z_vs_x1c} using all Pantheon and volume-limited Pantheon samples.

In the figure, we split the sample into high- and low-$x_1$ SNe Ia at $x_1 = -0.75$, empirically driven by an apparent drop in the observed $x_1$ distribution of the Pantheon sample.
However, this cut value defines the optimal boundary for creating a high-purity ($>93\%$) sample of low-$x_1$ SNe Ia, as demonstrated by a two-component Gaussian mixture model in the right sub-panel of Fig.~\ref{fig:z_vs_x1c} (see also \citealt{Wojtak2025}, who restricted their sample to $x_1 < -0.8$ to represent a high-purity low-stretch peak sample).
For the colour, we split the sample into red and blue SNe Ia based on the nominal value of $c = 0.0$.

SNe Ia with high-$x_1$ (young progenitor) are observed across the entire redshift range of the sample.
There are some low-$x_1$ (old progenitor) SNe Ia above $z = 0.5$, the fraction of them, however, starts to increase at $z \le 0.5$, as we expected.
Above that redshift, as presented in Tab.~\ref{tab:z_x1c}, only $12.5\%$ ($2.5\%$ for the volume-limited Pantheon sample) of SNe have low-$x_1$ value, which is compared to $23.2\%$ ($26.6\%$) at $z \le 0.5$. 
For $c$, the fraction of red SNe Ia shows little change between those at $z > 0.5$ and at $z \le 0.5$ ($35.6\%$ versus $38.1\%$ and $47.5\%$ versus $43.5\%$ for all and volume-limited Pantheon samples, respectively; see Tab.~\ref{tab:z_x1c}).

One might argue that this result, specifically the change in the fraction of high- to low-$x_1$ SNe Ia, is due to the Malmquist bias in the surveys, because $x_1$ is directly related to the intrinsic brightness of SNe Ia.
However, by examining the $c$ distribution, we can rule this out.
Since red SNe Ia are also observationally faint, the fraction of red SNe Ia should exhibit a decrease even more dramatic than that seen in low-$x_1$ because of the larger luminosity coefficient of colour than that of stretch (e.g., $\beta=3.0$ versus $\alpha=0.15$; \citealt{Scolnic2018}), if the Malmquist bias were the primary driver of the observed distribution at $z > 0.5$.
On the contrary, the fraction of red SNe Ia shows no significant change across $z\sim0.5$ ($43.5\%$ at $z\le0.5$ versus $47.5\%$ at $z>0.5$ for the volume-limited Pantheon sample), in stark contrast to the low-$x_1$ fraction ($26.6\%$ versus $2.5\%$).
The observation of these faint, red SNe Ia at high redshifts demonstrates that the surveys are sufficiently deep to detect them\footnote{We note that the absence of extreme red SNe Ia ($c > 0.3$) is a consequence of the colour cut applied to the Pantheon sample ($-0.3 < c < 0.3$; see \citealt{Scolnic2018}).}.
Furthermore, our empirical mapping of $x_1$ distribution based on SPAD predicts this change in $x_1$ distribution: the fraction of low-$x_1$ SNe Ia decreases as redshift increases (Fig.~\ref{fig:pred_x1}).
Therefore, we argue that the absence or less frequent observations of low-$x_1$ SNe Ia at high redshifts is not a selection effect, but an intrinsic property of SNe Ia, reflecting the host galaxy evolution with redshift, such that young galaxies and their corresponding young progenitor SNe Ia that have high-$x_1$ values dominate at high redshifts.
To fully confirm our finding, further SNe Ia data at high redshifts (e.g., $z > 0.5$) from a volume-limited survey are required.

\subsection{Creating a homogeneous sample of young SNe Ia from their observables across the entire redshift range}
\label{sec:homo_sneia}

\citetalias{Childress2014} suggested that SNe Ia selected only from actively star-forming galaxies would yield the most cosmologically uniform sample, due to the homogeneity of ``young'' SN Ia progenitor ages at all cosmological epochs.
Many previous studies support this by presenting a low rms scatter of the Hubble residual in those environments, from 0.065 mag to 0.172 mag depending on the sample used \citep[e.g.,][]{Kelly2015, Kim2018, Kim2024a}. 
To obtain such a homogeneous sample, additional host information is required\footnote{We note other studies that have proposed creating a homogeneous sample from various approaches using different host environments. \citet{Kelsey2023} found that blue SNe Ia in locally ($U - R$) blue environments have the lowest rms scatter of $0.141 \pm 0.016$ mag. However, as found in our work, the SN Ia colour distribution does not depend on the progenitor star age. Although the local blue colour may help, further studies are required to determine if this approach can produce a homogeneous young SN Ia sample. Another recent study by \citet{Ramaiya2025} found that massive and passive galaxies host a more uniform sample of SNe Ia, with the rms scatter of $0.151 \pm 0.025$ mag and also the shallowest colour-luminosity slope. However, as described below, $41.2\%$ of SNe Ia in locally passive environments have high-$x_1$ values. This demonstrates that while selecting massive and passive hosts might minimize dust-induced scatter, the sample still suffers from intrinsic heterogeneity across different $x_1$ populations, namely, having diverse progenitor star ages.}.
However, estimating the SN Ia local environments at $z>0.1$ is indeed challenging.
Instead, when we use the light-curve shape information as the indicator of the SN Ia progenitor star age, as proposed in the present work, it would be feasible to make such a homogeneous SN sample across the entire redshift range.
Moreover, for creating a more homogeneous sample, this light-curve shape-based selection is less sensitive to the Malmquist bias, as discussed above.
Since this sample is likely to be homogeneous in terms of the progenitor star age and also the SN Ia luminosity, there is also no need to worry about the possible evolution of SN Ia luminosity discussed in previous studies \citep[e.g.,][]{Kim2018, Kim2019, Kang2020, Son2025}.

This homogeneous SN Ia sample across the entire redshift range dominated by young progenitor stars can be primarily formed by selecting SNe Ia with a high-$x_1$ value.
However, one challenge is to define a cut value for $x_1$.
Based on the current data set of the Pantheon with the two-component Gaussian mixture model we analyzed (see Figure~\ref{fig:z_vs_x1c}), the high-$x_1$ value might be defined as $x_1 > -0.75$.
More quantitatively, \citetalias{Nicolas2021} and \citet{Ginolin2025a} determined means ($\mu_1$) and standard deviations ($\sigma_1$) of the high-$x_1$ SN Ia sample based on Eq.~\ref{eq:n21_x1} with the \citetalias{Nicolas2021} dataset and the ZTF SNIa DR2, respectively.
\citetalias{Nicolas2021} obtained $\mu_1 = 0.37 \pm 0.05$ and $\sigma_1 = 0.61 \pm 0.04$, and \citet{Ginolin2025a} obtained $\mu_1 = 0.42 \pm 0.08$ and $\sigma_1 = 0.54 \pm 0.05$, which show a good agreement with each other.
We can construct a homogeneous sample based on these values.

However, since we observe that the old progenitor stars (in locally old or passive environments) have both low- and high-$x_1$ SNe Ia, as shown in the left panel of Fig.~\ref{fig:localage}, we expect contamination by SNe Ia from old progenitors when only based on the $x_1$ value.
This contamination will begin to appear at $z \le 0.5$, where an old progenitor star and the passive sequence of galaxies begin to appear (see Fig.~\ref{fig:spad_z} and Fig.~\ref{fig:z_vs_x1c}, respectively), and will be the most abundant at the lowest redshift range.
For example, based on the \citet{Rigault2020} sample, which has accurately determined spectroscopic local star formation rate around an SN Ia explosion site, the contamination at $z < 0.1$ will be up to 41.2\% (42 high-$x_1$ SNe Ia in locally passive environments out of 102 high-$x_1$ SNe Ia).
Therefore, in order to select a correct cut criterion, a precise quantitative study of which value of $x_1$ is associated with which value of the progenitor star age is required.
This will be achieved by detailed analysis of the SN Ia explosion models with various progenitor properties and explosion mechanisms \citep[e.g.,][]{Leung2018, Leung2020, Gronow2021}, followed by observational confirmation \citep[e.g.,][]{Kim2025}.

\subsection{Future works}
\label{subsec:future_works}

We show in the present work that the SN Ia light-curve shape parameter would indicate the age of the progenitor star.
In other words, the progenitor star age, a strong candidate for the origin of the environmental dependence of SN Ia luminosities \citep[e.g.,][]{Sullivan2010, Pan2014, Kang2016, Kim2018, Kim2019, Kang2020, Chung2023}, therefore has an impact on the light-curve shape and thus the peak luminosity of SNe Ia.
However, we observe a wider dispersion around the observed peak of $x_1$ distribution than SPAD (see Fig.~\ref{fig:x1_spad}).
This indicates that the progenitor star age alone cannot fully account for the observed dispersion in $x_1$.
Several additional physical parameters are known to affect the peak luminosity.
\citet{Timmes2003}, for example, explored the impact of the progenitor metallicity on the $^{56}$Ni mass synthesized during the explosion, which is tightly related to the peak luminosity of SNe Ia.
They suggested that the progenitor metallicity could account for a 25\% variation in $^{56}$Ni mass, and thus 0.2 mag in the observed peak luminosity in the $V$-band.
The different SN Ia explosion mechanisms, such as near-Chandrasekhar or sub-Chandrasekhar mass and deflagration or detonation explosions, also contribute to the dispersion of the peak luminosity \citep[e.g.,][]{Leung2018, Leung2020}.
We expect that these combined effects can produce the broader distribution of the observed light-curve shape parameter, though a quantitative study using SN Ia explosion models is left to future work.

Regarding the colour of SNe Ia, the stellar colour is known to be related to its age.
However, extracting the intrinsic SN Ia colour is challenging because the light-curve colour parameter is a mixture of the intrinsic SN Ia colour and the dust around SNe Ia.
In the present work, we attempted to select normal SNe Ia by applying the colour cut, such as $|c|<0.3$.
However, as investigated by \citet{Ginolin2025b}, an SN Ia with $c > 0.2$ (or $0.1$ from \citealt{Brout2021}) is dominated by interstellar dust of host galaxies.
In the future, isolating purely dustless SNe Ia \citep[e.g.,][]{Ginolin2025b} or targeting SNe Ia that have a minimized colour-luminosity slope \citep[e.g.,][]{Chen2022, Ramaiya2025} will be essential to test whether a correlation between the intrinsic colour of SNe Ia and SPAD exists, analogous to that between the light-curve shape parameter and SPAD probed in the present work.

In addition, the community would benefit from an SN Ia light-curve fitter that has been trained on a more homogeneous sample, such as SNe Ia preferentially arising from young progenitor stars (i.e. high-$x_1$ SNe Ia).
As discussed in several studies \citep[e.g.,][]{Sullivan2010, Sullivan2011, Kim2018, Kim2019, Rigault2020, Ramaiya2025}, the absolute brightness of SNe Ia systematically differs depending on their host environments, ranging from $\sim$0.05 mag to $\sim$0.13 mag.
However, currently employed light-curve fitters are trained on heterogeneous samples that span the full range of host environments and progenitor star ages, effectively averaging over these distinct populations.
As a result, the standardization process from such training may introduce subtle systematic biases on determining cosmological parameters.
There are some efforts to analyze them separately.
For example, \citet{Ginolin2025a} introduced a broken-$\alpha$ model, employing two different slopes for high- and low-stretch SNe Ia.
In addition, as we presented in Sec.~\ref{sec:intro}, \citet{Wojtak2023} and \citet{Wojtak2025} introduced the two-population Bayesian hierarchical model and showed the possibility of mitigating the Hubble tension, providing $H_0$ = $70.59 \pm 1.15 \text{ km s}^{-1} \text{ Mpc}^{-1}$.
While developing a new fitter is beyond the scope of this paper, making separate fitters trained on different populations of SNe Ia and using a homogeneous sample of SNe Ia would be a next step for the community to yield more accurate distance estimates from SNe Ia in the forthcoming Rubin era \citep{Lsst2009}.

\begin{acknowledgements}
We thank the anonymous referee for the careful reading of the manuscript and for many constructive suggestions, which have significantly improved the clarity of this work. 

We acknowledge support from the National Research Foundation of Korea to the Center for Galaxy Evolution Research (RS-2022-NR070872, RS-2022-NR070525).

Y.-L.K. was supported by the Lee Wonchul Fellowship, funded through the BK21 Fostering Outstanding Universities for Research (FOUR) Program (grant No. 4120200513819) and the National Research Foundation of Korea grant funded by the Korea government(MSIT) (RS-2026-25473561).

S.-J.Y. acknowledges support from the Mid-career Researcher Program (RS-2024-00344283) through Korea's NRF funded by the Ministry of Science and ICT. 

This work used \textsc{\texttt{pandas}} \citep{McKinney2010}, \textsc{\texttt{numpy}} \citep{Harris2020}, and \textsc{\texttt{matplotlib}} \citep{Hunter2007}.
We also use the LINMIX package \href{https://github.com/jmeyers314/linmix/}{https://github.com/jmeyers314/linmix/}.

\end{acknowledgements}

\end{document}